\documentclass[
superscriptaddress,
aps,
prx,
reprint,
longbibliography
]{revtex4-2}

\usepackage{lineno}
\usepackage{graphicx}
\usepackage{setspace}
\usepackage{mathrsfs,dsfont,mathtools,bm, braket, amsmath, amsfonts}
\usepackage[dvipsnames]{xcolor}
\usepackage[
    colorlinks,
    linkcolor={blue!80!black},
    citecolor={blue!80!black},
    urlcolor={blue!80!black}
]{hyperref}

\usepackage[babel,kerning=true,spacing=true]{microtype}
\usepackage{verbatim}
\usepackage{soul}

\usepackage{multirow}
\usepackage{makecell}
\newcommand{\be}{\begin{equation}}
\newcommand{\ee}{\end{equation}}
\newcommand{\bea}{\begin{eqnarray}}
\newcommand{\eea}{\end{eqnarray}}

\newcommand{\lp}{\left(}
\newcommand{\rp}{\right)}

\def \cE{{\cal E}}

\def \cA{{\cal A}}

\def \cH{{\cal H}}

\usepackage{xcolor}

\usepackage[normalem]{ulem} 

\usepackage{amssymb}

\begin{document}


\title{Interaction driven charge transfer transitions in closely spaced graphene double layers}

\author{Kenneth A. Lin}
\thanks{These authors contributed equally to this work.}
\affiliation{Microelectronics Research Center, Department of Electrical and Computer Engineering,
The University of Texas at Austin, Austin, Texas 78758, USA}

\author{Unmesh Ghorai}
\thanks{These authors contributed equally to this work.}
\affiliation{School of Physics and Astronomy, Tel Aviv University, Tel Aviv 69978, Israel}

\author{Kenji Watanabe}
\affiliation{Research Center for Electronic and Optical Materials,
National Institute for Materials Science, 1-1 Namiki, Tsukuba 305-0044, Japan}

\author{Takashi Taniguchi}
\affiliation{Research Center for Materials Nanoarchitectonics,
National Institute for Materials Science, 1-1 Namiki, Tsukuba 305-0044, Japan}

\author{Emanuel Tutuc}
\email{etutuc@mail.utexas.edu}
\affiliation{Microelectronics Research Center, Department of Electrical and Computer Engineering,
The University of Texas at Austin, Austin, Texas 78758, USA}

\author{Rafi Bistritzer}
\email{rafib@tauex.tau.ac.il}
\affiliation{School of Physics and Astronomy, Tel Aviv University, Tel Aviv 69978, Israel}
\vspace{1cm}

\begin{abstract}
\vspace{1em}
\begin{doublespace}
\bfseries
Charge transfer between two conductors is conventionally viewed as a single-particle process governed by electrostatics and band alignment. Using tunneling spectroscopy, we show that charge transfer in closely spaced graphene double layer quantum Hall ferromagnets instead proceeds through a sequence of interaction driven phase transitions governed by the competition between capacitive charging and Coulomb exchange interactions. A comparison of experimental data and theoretical calculations identifies spectroscopic signatures of the interaction driven charge transfer transitions, and reveals that this charge transfer reconstructs the quasiparticle spectrum. While intralayer exchange favors abrupt transfer of entire spin-valley subbands between the layers, interlayer exchange stabilizes coherent intermediate phases that enable gradual charge transfer. Our results establish interlayer tunneling as a powerful probe of interacting electronic systems whose quasiparticle spectrum is itself bias dependent.
\end{doublespace}
\end{abstract}

\maketitle

\onecolumngrid
\doublespacing

Strong electronic interactions can drive collective quantum phases when kinetic energy is suppressed. Recent examples include moir\'e materials and rhombohedral graphene multilayers, where narrow electronic bands have enabled correlated insulators \cite{YCaoTBGCorrelatedStates2018, WillBurgCorrelatedStatesTDBG2019, ShenCorrelatedStatesTDBG2020, XiaomengCorrelatedStatesTDBG2020, WangCorrlatedStatesTwistedTMD2020, WillBurgATQGCorrelations2022, HanCorrelatedInsulatorChernInsulatorRhombohedral2024}, superconductivity \cite{YCaoSuperconductivityMATBG2018, JaneParkSuperconductivityMATTG2021, ZhouSuperconductivityRhombohedral2021, JaneParkMultilayerGrapheneMagicSuperconductivity2022, YZhangMultilayerGrapheneSuperconductivity2022, KinFaiMakTwistedWSe2Supercon2025, CoryDeanTwistedWSe2Supercon2025}, magnetism \cite{AaronSharpeFerromagnetismTBG2019, ZhouFerromagnetismRhombohedral2021}, and topological states \cite{SerlinQAHEMoire2020, NuckollsChernMATBG2020,  XieChernInsulatorMATBG2021, WuChernMATBG2021, YimengBulkEdgeTDBG2022, HanCorrelatedInsulatorChernInsulatorRhombohedral2024}. The extreme limit of interaction-dominated physics is realized in Landau levels (LLs), where kinetic energy is quenched and electrons occupy macroscopically degenerate quantum states. This limit underlies the fractional quantum Hall effect \cite{TsuiFracQuanHallEffect1982, LaughlinFracQuantHallEff1983, highB_th_GMP_1986, PKimFractQHEinGraphene2009, CoryDeanGrapheneFracQHE2011}, Wigner crystals \cite{GrimesWignerCrystal1979, EYAndreiWignerCrystal1988, SantosWignerCrystal1992}, and the stripe and bubble phases observed in higher LLs \cite{MoessnerStripeBubble1996, PanStripeBubble1999, LillyStripeBubble1999, StanescuStripeBubble2000}. In multicomponent systems with additional degrees of freedom, exchange interactions can furthermore lift spin, valley, or orbital degeneracies and produce quantum Hall ferromagnets with a rich variety of ordered states 
\cite{highB_th_Fertig_spectrum_layered_highB_1989,
quant_ferro_qhs_Yang_1994, int_coh_qhs_merons_Moon_1995, review_Girvin_MacDonald_multicomponent_QH_1995, SpielmanResonantlyEnhancedDoubleLayerQuantumHallFerromagnet2000, qhf_collec_mode_Spielman_2001, qhf_exp_Tutuc_density_imbalance_WC_2003, qhf_exp_Spielman_density_imbalance_2004, qhf_exp_Champagne_charge_imbalance_2008, exp_Young_gs_gv_2012, exp_Jiang_graphene_gs_gv_2019}. 

Double layer systems provide an additional layer degree of freedom that can be controlled by an interlayer bias. In double layer systems with interlayer spacing $\approx$ 10 nm or larger, such as GaAs/AlGaAs heterostructures, electrostatic charging significantly constrains charge redistribution between the layers. Charge transfer has nevertheless been observed in narrowly tuned regimes: at zero magnetic field near a particular balance between interlayer tunneling and Coulomb interactions set by the barrier thickness \cite{chargeTranfer_Katayama_1995}, and in the fractional quantum Hall regime at specific fillings where unequal layer populations stabilize incompressible states \cite{chargeTransfer_Manoharan_1997}. 

Closely spaced graphene double layers subjected to strong magnetic fields \cite{XiaomengQuantumHallDragExcitonGraphene2017, JIALiExcitonSuperfluidDoubleBilayerGraphene2017, EmergenceOfInterlayerCoherenceInKenneth2022} 
establish a qualitatively different regime where atomic-scale layer separation makes capacitive charging, intralayer exchange, and interlayer exchange comparable, while bare tunneling remains very small. Here, by utilizing interlayer tunneling, we show that charge transfer in closely spaced graphene double layers is a many-body phenomenon governed by the competition between electrostatic charging and exchange interactions.  In particular, intralayer exchange can overcome the capacitive cost of charge imbalance, destabilizing states with partially filled LL subbands. This instability is analogous to itinerant ferromagnetism: just as exchange interactions favor spin polarization in a metal, intralayer exchange favors layer polarization by lowering the energy when electrons concentrate within fewer spin-valley-resolved LLs. Consequently, charge transfer no longer occurs continuously, but instead proceeds through collective rearrangements between the layers, giving rise to a sequence of charge transfer transitions between competing quantum Hall ferromagnetic configurations. Interlayer exchange partially offsets this tendency by stabilizing coherent superpositions of the two layers, which bridge neighboring layer-polarized states through continuous quantum phase transitions. 

Our results further reveal a regime in which electrical bias drives a reconstruction of the interacting electronic spectrum rather than merely shifting the relative energies of two conductors. Accordingly, our observations point to a new tunneling paradigm where interlayer tunneling probes the many-body evolution of a single strongly interacting system. This stands in stark contrast to conventional frameworks, where tunneling measures the spectral functions of two nearly independent conductors, or couples directly to the interlayer order parameter in coherent quantum Hall bilayers.

\section*{Graphene double layer heterostructure}

\begin{figure*}
\center\includegraphics[width=\textwidth]{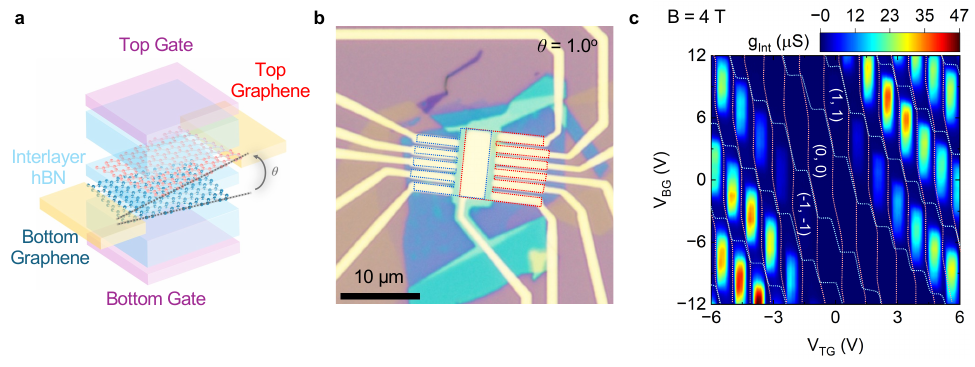}
\caption{\label{fig:Fig1}
\textbf{Graphene double layer and tunneling in the quantum Hall regime. }
\textbf{a,} Schematic of the dual-gated graphene double layer heterostructure consisting of two independently contacted graphene monolayers separated by a four-layer hBN tunnel barrier. The graphene layers have a relative twist angle $\theta = 1.0^{\circ}$.
\textbf{b,} Optical micrograph of the graphene double layer heterostructure. The independently contacted top (bottom) graphene electrodes are outlined in red (blue). 
\textbf{c,} $g_{\mathrm{Int}} = dI_{\mathrm{Int}}/dV_{\mathrm{Int}}$ versus $V_{\mathrm{TG}}$ and $V_{\mathrm{BG}}$ at $V_{\mathrm{Int}} = 0$, $B = 4$ T and $T = 1.5$ K. The red (blue) dashed lines indicate boundaries of the orbital Landau level (LL) sectors of the top (bottom) graphene layer. The orbital LL sectors $(N_{\mathrm{T}}, N_{\mathrm{B}}) = (0, 0)$ and $(\pm1, \pm1)$ are marked. 
}
\end{figure*}

The double layer heterostructure we study consists of two independently contacted graphene monolayers separated by a four-layer thick hexagonal boron-nitride (hBN) tunnel barrier, corresponding to an interlayer separation of $d = 1.8$ nm (Fig. \ref{fig:Fig1}a). The thin tunnel barrier places the system in the exchange-dominated regime, $d/\ell_B  \lesssim 1$, where $\ell_B = \sqrt{\hbar / eB}$ is the magnetic length. The heterostructure is encapsulated in additional top and bottom hBN dielectrics \cite{hBNSubstrates2010}, with top and bottom gates that independently control the carrier densities in the two graphene layers. The top gate ($C_{\mathrm{TG}}$), interlayer ($C_{\mathrm{Int}}$), and bottom gate ($C_{\mathrm{BG}}$) capacitances are 74 nF/cm$^2$, 1.7 $\mu$F/cm$^2$, and 8.1 nF/cm$^2$ respectively. The relative  twist angle between the two graphene layers is $\theta=1^{\circ}$, determined from the resonant tunneling characteristics at zero magnetic field (see Appendix A). The small twist in sample design is chosen in order to suppress the high tunneling conductance peaks associated with commensurate angles \cite{transportTBG, EmergenceOfInterlayerCoherenceInKenneth2022}. An optical micrograph of the device studied is shown in Fig. \ref{fig:Fig1}b. 

To perform tunneling measurements, we apply an interlayer voltage on the top layer while keeping the bottom layer at ground. Multiple contacts to the two graphene layers enable simultaneous measurement of the tunneling current ($I_{\mathrm{Int}}$) and the four-probe interlayer voltage ($V_{\mathrm{Int}}$). Figure \ref{fig:Fig1}c shows tunneling conductance ($g_{\mathrm{Int}} = dI_{\mathrm{Int}}/dV_{\mathrm{Int}}$) in the limit of zero interlayer bias, as a function of top ($V_{\mathrm{TG}}$) and bottom ($V_{\mathrm{BG}}$) gate voltages, at temperature $T = 1.5$ K and perpendicular magnetic field $B = 4$ T. Clear oscillations in the $g_{\mathrm{Int}}$ data are associated with the LLs of both layers, where maxima (minima) denote partial (full) filling of the orbital LL sectors \cite{EmergenceOfInterlayerCoherenceInKenneth2022}. In this work, we focus on the regime in which the chemical potentials of the top ($\mu_{\mathrm{T}}$) and bottom ($\mu_{\mathrm{B}}$) layers lie within the lowest orbital LL manifold $(N_{\mathrm{T}}, N_{\mathrm{B}}) = (0, 0)$, as highlighted in Fig. \ref{fig:Fig1}c.

Each graphene layer hosts four spin-valley subbands corresponding to the two spin states $s=\uparrow,\downarrow$ and two valleys $\tau=K,K'$. The perpendicular magnetic field lifts the spin-valley degeneracy by the Zeeman $\Delta_Z=g_s\mu_{\mathcal{B}} B$ and valley $\Delta_V=g_v\mu_{\mathcal{B}} B$ splittings; $g_s$ and $g_v$ are the spin and valley g-factors, respectively, and $\mu_{\mathcal{B}}$ is the Bohr magneton. The single-particle energies are thus $\epsilon_i=\pm\Delta_Z/2 \pm \Delta_V/2$, where $i=(\tau,s)$ runs over the four spin-valley subbands $K\downarrow, K\uparrow, K'\downarrow, K'\uparrow$ in order of increasing energy. Throughout this work, we consider the $\nu=0$ state, $[\nu_ {\mathrm{T}}=0,\nu_ {\mathrm{B}}=0]$, as the zero-bias reference configuration, in which the two lowest spin-valley subbands in the orbital $N=0$ LL of each layer are filled while the two upper subbands are empty (we discuss the $\nu = 0$ state with nonzero charge imbalance $\nu_{\mathrm{T}} = -\nu_{\mathrm{B}}$ in Appendix B). Because the interlayer capacitance is much larger than the top- and bottom-gate capacitances, the total filling factor $\nu=\nu_ {\mathrm{T}}+\nu_ {\mathrm{B}}$ remains, to a good approximation, fixed at the value set by the gate voltages, while the interlayer bias drives charge transfer between the two layers by setting the chemical potential difference $\mu_ {\mathrm{T}}-\mu_ {\mathrm{B}} = eV_{\mathrm{Int}}$. Because the bare interlayer tunneling energy is $w \simeq 1-10 \mu{\rm eV}$ \cite{KennethTunnelingEnergyhBNBarrier2025}, several orders of magnitude smaller than the Zeeman, valley, and interaction energy scales, tunneling is neglected when determining the equilibrium occupations and the ground state. 

\section*{Interaction driven charge transfer} 

\begin{figure*}
\center\includegraphics[width=\textwidth]{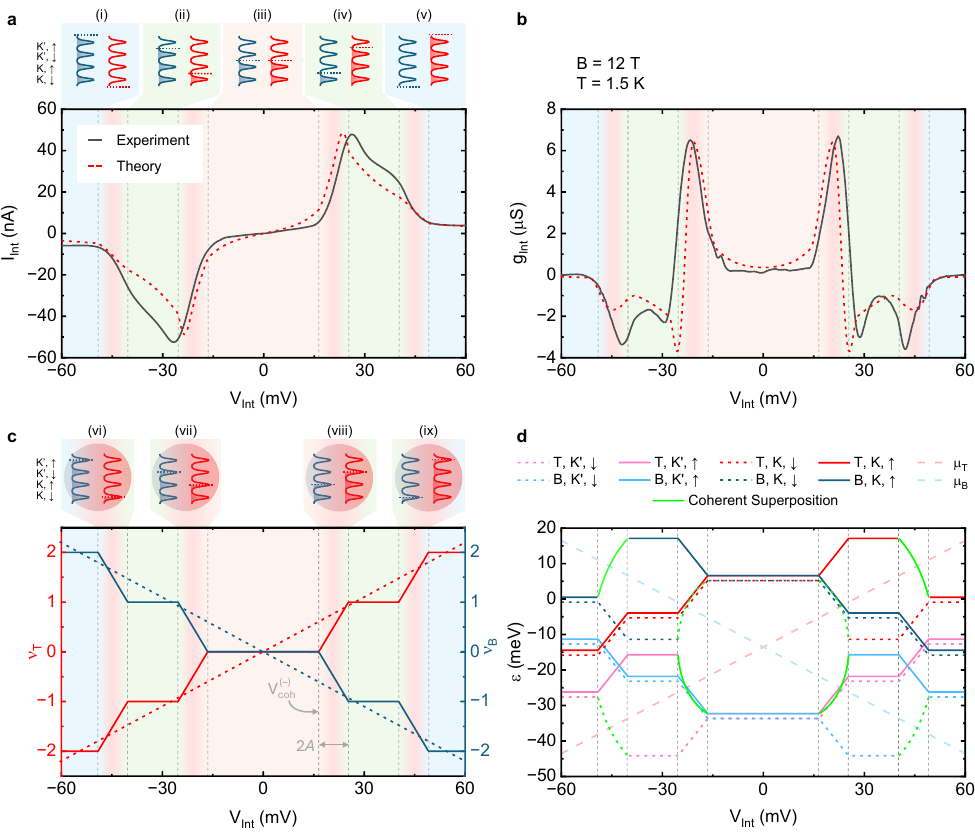}
\caption{\label{fig:Fig2}
\textbf{Tunneling spectroscopy of interaction driven charge transfer.}
\textbf{a--b,} Measured (solid) and calculated (dashed) interlayer tunneling current $I_{\mathrm{Int}}$ and differential conductance $g_{\mathrm{Int}}=dI_{\mathrm{Int}}/dV_{\mathrm{Int}}$ versus interlayer bias $V_{\mathrm{Int}}$ at $B=12$ T and $T=1.5$ K. At the bias-driven phase transitions, the reconstruction of the quasiparticle spectrum
gives rise to kinks in $I_{\mathrm{Int}}$ and corresponding extrema in $g_{\mathrm{Int}}$. The agreement between theory and experiment demonstrates that tunneling spectroscopy directly probes the interaction-reconstructed quasiparticle spectrum.
\textbf{c,} Calculated top (red) and bottom (blue) layer filling factors $\nu_{\mathrm{T}}$ and $\nu_{\mathrm{B}}$ as a function of $V_{\mathrm{Int}}$.
Solid lines show the interacting ground state. The dotted lines show the charge redistribution from an independently determined interlayer capacitance \cite{KennethTunnelingEnergyhBNBarrier2025}, neglecting Landau quantization and exchange interactions. Charge transfer proceeds through a sequence of subband-polarized quantum Hall ferromagnets connected by coherent intermediate phases (red shaded regions).
\textbf{d,} Interacting quasiparticle spectrum as a function of $V_{\mathrm{Int}}$. The green branches denote quasiparticle energies in the interlayer coherent phases, arising from hybridization of top- and bottom-layer spin-valley subbands. The dashed lines denote the chemical potentials of the top and bottom layers.
}
\end{figure*}

We now turn to the tunneling measurements at total filling factor $\nu=0$. Figure~\ref{fig:Fig2}a shows the measured tunneling current $I_{\mathrm{Int}}$ as a function of the interlayer bias $V_{\mathrm{Int}}$ (solid curve) at $T=1.5$ K and $B=12$ T, together with the corresponding differential conductance $g_{\mathrm{Int}}=dI_{\mathrm{Int}}/dV_{\mathrm{Int}}$ in Fig.~\ref{fig:Fig2}b. Rather than exhibiting a sequence of tunneling resonances, the current displays an unusual non-monotonic dependence on bias. The differential conductance remains strongly suppressed up to $V_{\mathrm{Int}}\approx \pm 20$ mV despite the much smaller Zeeman ($\approx$ 1 meV) and valley ($\approx$ 10 meV) splittings. 
At $V_{\mathrm{Int}} \approx \pm 20$ mV the current magnitude rises abruptly, before decreasing gradually through two pronounced kinks that appear as sharp negative peaks in $g_{\mathrm{Int}}$. These characteristic voltages do not coincide with any single-particle energy scale, suggesting that the tunneling spectrum is governed by interaction driven physics. 
As we show below, each of these features in the tunneling current marks a phase transition that reorganizes the equilibrium charge distribution between the two layers [Fig. \ref{fig:Fig2}a(i)--(v)].

To explain the microscopic origin of the tunneling features, we consider the equilibrium many-body ground state of the coupled graphene double layer within the framework of the Hartree-Fock approximation \cite{highB_th_Fertig_spectrum_layered_highB_1989, qhf_th_Joglekar_bias_induced_phasr_transition_2002}. Projected to the $N=0$ LL, the mean field energy
\begin{equation}
\mathcal{E}_{\rm HF} =
\sum_{\ell,i}\epsilon_i n_{\ell i}
+ \frac{v_x}{2}\left(\Delta\nu\right)^2
- \frac{\Gamma_A}{2}\sum_{\ell,i}n_{\ell i}^2
- \Gamma_E m^2
- \frac{e V_{\mathrm{Int}}}{2}\Delta\nu
\label{EMF}
\end{equation}
is determined by the occupations $n_{\ell i}$ of the eight spin-valley subbands and by the interlayer coherence $m$. Here $\ell=\mathrm T, \mathrm B$ labels the layer, $\nu_\ell=\sum_i n_{\ell i} - 2$ are the layer filling factors, and $\Delta\nu=\nu_\mathrm{T}-\nu_\mathrm{B}$ is the layer polarization. The second term is the electrostatic charging energy, where $v_x = \frac{d}{2 \ell_B} \frac{\epsilon_\parallel}{\epsilon_\perp} E_C$ and $E_C = \frac{e^2}{\epsilon_\parallel l_B}$ is the Coulomb energy scale. The third and fourth terms are, respectively, the intralayer and interlayer exchange energies, where $\Gamma_A = \sqrt{\frac{\pi}{2}} E_C$ and
$\Gamma_E = \Gamma_A \exp(\frac{a^2}{2}) \mathrm{erfc}(\frac{a}{\sqrt{2}})$, with $a = \frac{d}{l_B} \sqrt{\frac{\epsilon_\parallel}{\epsilon_\perp}}$. The final term couples the layer polarization to the applied interlayer bias. The different contributions to Eq.~(\ref{EMF}) favor distinct electronic configurations. The electrostatic charging energy penalizes layer polarization and therefore favors equal electron densities in the two graphene layers. The intralayer exchange energy favors layer polarization by lowering the energy when electrons occupy fewer spin-valley subbands. By contrast, the interlayer exchange energy is minimized when electrons occupy coherent superpositions of the two layers, allowing exchange to be gained without fully transferring charge between the layers. Finally, the applied interlayer bias lowers the energy of layer-polarized states with $\Delta\nu>0$, thereby driving charge transfer between the layers. For a fixed total filling factor $\nu=\nu_\mathrm T+\nu_\mathrm B$, the ground state is obtained by minimizing Eq.~(\ref{EMF}) with respect to the occupations $n_{\ell i}$ and the interlayer coherence $m$, subject to the constraints
$0\le n_{\ell i}\le1$, and $\sum_{\ell,i}n_{\ell i}=\nu+4$.
The coherence is further constrained by the fermionic structure of the one-body density matrix. For coherence between subbands $\mathrm T_i$ and $\mathrm B_j$,
$m^2 \le \min \left\{ n_{\mathrm T_i}n_{\mathrm B_j}, (1-n_{\mathrm T_i})(1-n_{\mathrm B_j})
\right\}$. Because the interlayer exchange energy decreases monotonically with $m^2$, the ground state always saturates this bound.

We now examine how the $\nu=0$ state evolves with $V_{\mathrm{Int}}$. We begin near $V_{\mathrm{Int}} = 0$ [Fig. \ref{fig:Fig2}a(iii), orange shaded], where electrons occupy the lower two subbands of both layers, that is $\nu_{\mathrm{T}} = \nu_{\mathrm{B}} = 0$. The essential physics of charge redistribution is captured by considering the first charge transfer transition, in which electrons are transferred from the second occupied subband of the bottom layer to the third subband of the top layer. The occupations are then parameterized by $n_{\mathrm T3}=x$, $n_{\mathrm B2}=1-x$, with $0\le x\le1$, while all remaining occupations retain their values in the $\Delta\nu=0$ state. The coherence satisfies $m^2=x(1-x)$. The resulting energy is $\mathcal{E}_{32}(x) = \mathcal{E}_{22} + \left( eV_{\rm coh}^{(-)} -e V_{\mathrm{Int}} \right)x + \mathcal{A}x^2$,
where $\mathcal{A}  = 2v_x-\Gamma_A+\Gamma_E$, $V_{\rm coh}^{(-)} = \epsilon_3-\epsilon_2+\Gamma_A-\Gamma_E$,
and $\mathcal{E}_{22} = 2(\epsilon_1+\epsilon_2-\Gamma_A)$.
The curvature $\mathcal{A}$ measures the competition between capacitive charging, intralayer exchange, and interlayer exchange.
The energy is always convex  ($\mathcal{A}>0$). Therefore, an energy minimum is obtained either at one of the fully polarized boundaries ($x=0$ or $x=1$) where $m=0$, or at an interior minimum ($0<x<1$) where the two subbands are coherent.
For $V_{\mathrm{Int}} < V_{\rm coh}^{(-)}$, the system lacks coherence ($x=0$) and remains pinned in the $\Delta \nu = 0$ state [Fig. \ref{fig:Fig2}a(iii) orange shaded]. For  $V_{\rm coh}^{(-)} < V_{\mathrm{Int}} < V_{\rm coh}^{(-)} + 2\cA$, the system is in a coherent state ($0<x<1$) and charge is transferred continuously through coherent superpositions of the two layers [Fig. \ref{fig:Fig2}c(viii), red shaded]. Finally, at 
$V_{\mathrm{Int}} > V_{\rm coh}^{(-)}+2\cA$, the entire bottom layer subband is transferred to the top layer ($x=1$) and $\Delta \nu=2$ is again pinned to an integer value, resulting in the layer fillings $\nu_{\mathrm{T}} = 1$ and $\nu_{\mathrm{B}} = -1$ [Fig. \ref{fig:Fig2}a(iv), green shaded]. 
Repeating the same argument for the next pair of spin-valley subbands yields a similar charge transfer transition which completes the transfer of the entire charge in the $N=0$ LL to the top layer (see Appendix D).  

Figure \ref{fig:Fig2}c shows the calculated ground state top (bottom) layer filling $\nu_{\mathrm{T}}$ ($\nu_{\mathrm{B}}$) versus $V_{\mathrm{Int}}$ (solid lines). The data show subband-polarized quantum Hall ferromagnetic regimes in which layer filling factors $\nu_{\mathrm{T}}$ and $\nu_{\mathrm{B}}$ are constant [Fig. \ref{fig:Fig2}a(i)--(v)], bridged by interlayer coherent intermediate states where charge is continuously redistributed between neighboring subband-polarized states [Fig. \ref{fig:Fig2}c(vi)--(ix)]. The step-like dependence of $\Delta \nu$ on $V_{\mathrm{Int}}$ is due to the small value of $d/\ell_B$ which narrows the voltage range of the coherent phases (see Appendix G).

In Fig. \ref{fig:Fig2}d, we show the calculated quasiparticle energies. The elementary excitations are determined by the eigenvalues of the Hartree-Fock Hamiltonian $H_{\rm HF} = \partial\mathcal{E}_{\rm MF}/\partial \rho$ where $\rho$ is the one-body density matrix and $\cE_{\rm MF} = \cH_{\rm HF} + \frac{e V_{\mathrm{Int}}}{2}\Delta\nu$. Within the subband-polarized phases where the coherence vanishes [red (blue) lines in Fig. \ref{fig:Fig2}d for top (bottom) layer], the quasiparticle energies are $\epsilon_{\ell,i} = \epsilon_i - \Gamma_A n_{\ell,i} \pm v_x \Delta \nu$, where the $+$ $(-)$ sign corresponds to the top (bottom) layer. In the interlayer coherent phases (green curves in Fig. \ref{fig:Fig2}d), however, interlayer exchange hybridizes the partially occupied subbands of the two layers. For example, for the coherent state in Fig. \ref{fig:Fig2}c(viii), the energies are $E_\pm = \bar{\epsilon} \pm \sqrt{\lp \delta \epsilon \rp^2 + \Gamma^2_E m^2}$ where $\bar{\epsilon} = (\epsilon_{\mathrm{T}3}+\epsilon_{\mathrm{B}2})/2$ and $\delta \epsilon = (\epsilon_{\mathrm{T}3}-\epsilon_{\mathrm{B}2})/2$. Near zero bias, the two lowest spin-valley subbands are occupied in each layer. As the bias increases, the system passes through two interlayer coherent phases that connect successive subband-polarized configurations, ultimately producing a fully layer-polarized state. The concomitant evolutions of the charge distribution and quasiparticle spectrum are shown in Fig.~\ref{fig:Fig2}c--d. As we will show, this quasiparticle spectrum reconstruction accounts for the unusual tunneling characteristics observed in Figs.~\ref{fig:Fig2}a--b.

\section*{Tunneling in the strong coupling regime} 

The reconstruction of the quasiparticle spectrum provides the microscopic basis for the tunneling characteristics. We calculate the tunneling current perturbatively to leading order in the weak tunneling amplitude $w$.  The resulting current (see Appendix F)
\begin{equation}
I(V) = C \int_{\bar{\mu}-eV}^{\bar{\mu}} \frac{d\omega}{2\pi}
\sum_i A_{\mathrm Ti}(\omega; V_{\mathrm{Int}})
A_{\mathrm Bi}(\omega+eV_{\mathrm{Int}}; V_{\mathrm{Int}}),
\label{eq:Current}
\end{equation}
is a sum over the four spin-valley flavors reflecting their conservation in the tunneling process \cite{transportTBG, moireBands}. Here $\bar{\mu}$ is the chemical potential of the tunneling-free interacting double layer fixed by the total filling constraint. The prefactor $C \propto  w^2 \exp\!\left( -k_\theta^2\ell_B^2/2 \right)$ is exponentially suppressed by $k_\theta \approx k_{\rm D} \theta$ with $k_{\rm D}$ being the Dirac momentum. The twist angle therefore controls only the magnitude of the tunneling current, while the voltages at which the charge transfer transitions occur are independent of $\theta$. 
To facilitate comparison with experiment, the disorder broadening is approximated by a constant, allowing Eq.~(\ref{eq:Current}) to be evaluated analytically (see Appendix F,3).
Although Eq.~(\ref{eq:Current}) retains the convolution form of conventional tunneling theory, its physical content is fundamentally different because the spectral functions $A_{\ell i}(\omega; V_{\mathrm{Int}})$ evolve with the interacting ground state at each applied bias, thereby determining both the positions and the lineshapes of the tunneling features.

The dashed lines in Fig. \ref{fig:Fig2}a--b show the calculated $I_{\mathrm{Int}}$ and $g_{\mathrm{Int}}$ versus $V_{\mathrm{Int}}$ at $B = 12$ T for the $\nu=0$ state. The calculated curves reproduce the measured tunneling characteristics remarkably well. Within each subband-polarized phase, the spectral functions consist of single broadened quasiparticle peaks centered at the Hartree-Fock quasiparticle energies, and the current evolves smoothly with $V_{\mathrm{Int}}$. At the charge transfer transitions, each coherent subband pair contributes two broadened quasiparticle peaks whose spectral weights are determined by the coherence factors of the interacting eigenstates (see Appendix F,4), producing extrema  in $g_{\mathrm{Int}}$, and the corresponding kinks in $I_{\mathrm{Int}}$. 
The observed tunneling features therefore originate from interaction driven charge transfer rather than from tunneling between two rigid electronic spectra. Interlayer tunneling spectroscopy thus provides a direct experimental probe of interaction driven charge transfer and the accompanying reconstruction of the quasiparticle spectrum.

\section*{Magnetic field and temperature dependence} 

\begin{figure*}
\center\includegraphics[width=0.8\textwidth]{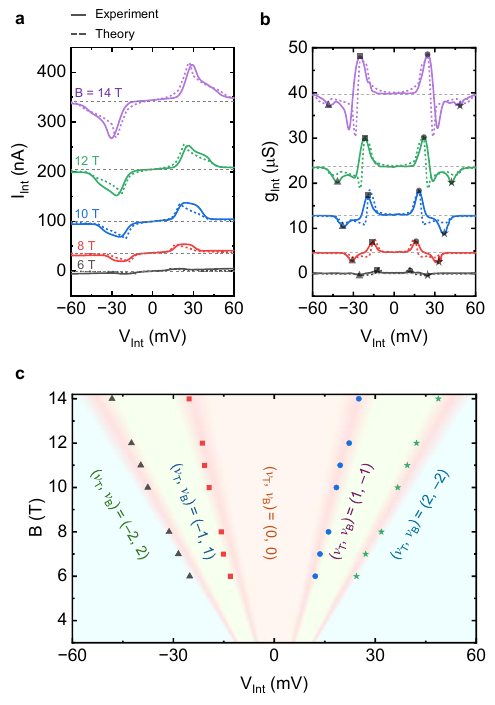}
\caption{\label{fig:Fig3}
\textbf{Magnetic field evolution of interaction driven charge transfer.}
\textbf{a--b,} Measured (solid) and calculated (dashed) $I_{\mathrm{Int}}$ and $g_{\mathrm{Int}}$ versus $V_{\mathrm{Int}}$ for magnetic fields between $B=6$ and $14$ T at $T = 1.5$ K. The curves are offset for clarity. The positions of the conductance peaks, marked by symbols, track the evolution of the interaction driven charge transfer transitions with magnetic field.
\textbf{c,} Phase diagram mapping the calculated top and bottom layer filling factor $(\nu_{\mathrm{T}}, \nu_{\mathrm{B}})$ as functions of $V_{\mathrm{Int}}$ and $B$. The red shaded regions indicate the coherent phases connecting subband-polarized quantum Hall ferromagnets. The symbols mark the conductance peak positions from panel \textbf{b} and coincide with the calculated centers of the coherent phases. The agreement between the measured and calculated transition voltages over the full magnetic field range demonstrates that tunneling spectroscopy tracks the evolution of the interacting double layer ground state.}
\end{figure*}

Figure \ref{fig:Fig3}a compares the measured $I_{\mathrm{Int}}$ versus $V_{\mathrm{Int}}$ at $T = 1.5$ K for magnetic fields $B = 6$ to $14$ T (solid lines), and the theoretical curves (dashed lines). The corresponding differential conductance $g_{\mathrm{Int}}$ is shown in Fig. \ref{fig:Fig3}b, with symbols identifying the centers of the four charge transfer transitions. The magnetic field evolution of these tunneling features is summarized in Fig. \ref{fig:Fig3}c. 
The lower bias features (circles in Fig. \ref{fig:Fig3}b--c) correspond to the center of the charge transfer transition between the $(\nu_{\mathrm{T}}, \nu_{\mathrm{B}}) = (0, 0)$ and  $(\nu_{\mathrm{T}}, \nu_{\mathrm{B}}) = (1, -1)$ configurations, while the higher bias features (stars in Fig. \ref{fig:Fig3}b--c) correspond to the center of the charge transfer transitions between the $(1, -1)$ and  $(2, -2)$ configurations. The theoretical transition voltages depend linearly on magnetic field as $eV_{(\nu_{\mathrm{T}}, \nu_{\mathrm{B}})=(0,0)\rightarrow(1,-1)} = \mu_B B \left( g_v-g_s+\frac{2d}{\varepsilon_\perp a_B} \right)$ and $eV_{(\nu_{\mathrm{T}}, \nu_{\mathrm{B}})=(1,-1)\rightarrow(2,-2)} = \mu_B B \left( g_v+g_s+\frac{6d}{\varepsilon_\perp a_B} \right)$ respectively, with matching features at negative $V_{\mathrm{Int}}$ related by the approximate particle-hole symmetry of the $\nu=0$ state (see Appendix C). 
Quantitative agreement between experiment and theory is obtained using  $g_s = 2$, $g_v = 17$, $\epsilon_\perp = 4.5$ and $\epsilon_\parallel = 9.0$. These values are similar to previously reported values for graphene \cite{exp_Young_gs_gv_2012, LyonGrapheneGS2017, EichGrapheneGV2018, LiSiYuGrapheneGV2019, BanszerusGrapheneGV2021} and hBN \cite{GeickhBNEpsilonParallel1966, YuEliashBNEpsilonParallel2013, LaturiahBNEpsilonParallel2018, LeehBNEpsilonParallelQTM2026}. The excellent agreement between the measured and predicted transition voltages over the full magnetic field range (Fig.~\ref{fig:Fig3}c)  confirms the interaction driven charge transfer mechanism.

\begin{figure*}
\center\includegraphics[width=\textwidth]{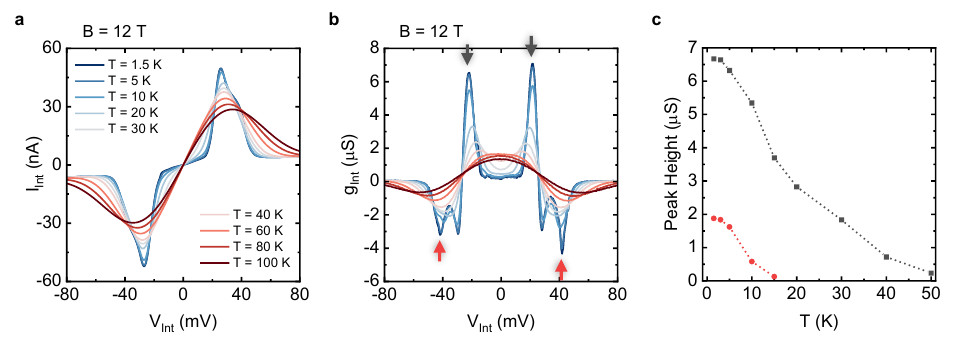}
\caption{\label{fig:Fig4}
\textbf{Temperature dependence of the tunneling characteristics.}
\textbf{a--b,} Measured $I_{\mathrm{Int}}$ and $g_{\mathrm{Int}}$ versus $V_{\mathrm{Int}}$ at $B=12$ T for temperatures between $T=1.5$ K and $100$ K. 
\textbf{c,} Peak heights of the symmetric conductance features marked by the gray and red arrows in \textbf{b}. As the temperature increases, the sharp
conductance peaks associated with the charge transfer transitions broaden and gradually disappear. The persistence of these tunneling signatures to
temperatures of several tens of kelvin reflects the large interaction energy scales driving the charge transfer transitions.}
\end{figure*}

We next examine the temperature dependence. In Fig. \ref{fig:Fig4}a--b, we show $I_{\mathrm{Int}}$ and $g_{\mathrm{Int}}$ as a function of $V_{\mathrm{Int}}$ at $B = 12$ T, for temperatures from $T = 1.5$ K to $100$ K. In Fig. \ref{fig:Fig4}c, we plot the peak height marking the charge transfer transitions as a function of temperature, relative to the tunneling background at the temperature when the signatures vanish. The signatures of the symmetric charge transfer transition between the $(0, 0)$ and  $(\pm1, \mp1)$ configurations (gray symbols) vanish at $T = 50$~K. In contrast, signatures of the symmetric charge transfer transition between the $(\pm1, \mp1)$ and  $(\pm2, \mp2)$ configurations (red symbols) vanish at $T = 20$ K. The earlier disappearance of the higher-bias features is likely associated with the larger quasiparticle broadening observed away from the Fermi level in graphene double layers \cite{NitinQuantumLifeSpectroscopy2021}.
The robustness of the tunneling signatures up to temperatures of several tens of kelvin reflects the large interaction energy scales driving the charge transfer transitions. 

\section*{Discussion} 

We have shown that charge transfer in closely spaced graphene double layer quantum Hall ferromagnets is fundamentally a many-body phenomenon governed by the competition between capacitive charging and Coulomb exchange interactions. Charge is redistributed through coherent intermediate phases that connect neighboring quantum Hall ferromagnets with different integer spin-valley occupations. Interlayer coherence is therefore not merely an additional ordered phase, but the mechanism that enables continuous charge transfer while retaining a substantial fraction of the exchange energy. 

A distinctive aspect of this system is the role played by tunneling spectroscopy. In a conventional tunnel junction, the applied bias probes the spectral functions of two nearly independent conductors. Here, by contrast, the interlayer bias also drives charge transfer and thereby changes the many-body ground state and quasiparticle spectrum of the coupled double layer. The abrupt changes and pronounced kinks in the $I(V)$ curve therefore do not reflect the alignment of two rigid Landau level spectra, but rather quasiparticle spectrum reconstruction as the system passes through successive charge transfer transitions. Tunneling spectroscopy thus probes the bias-driven evolution of a single strongly interacting double layer system, whose two layers cannot be treated as independent electronic subsystems.

More broadly, our results identify a regime in which capacitive charging, intralayer exchange, and interlayer exchange compete on comparable energy scales, allowing an applied electrical bias to drive collective many-body evolution. They establish interaction driven charge transfer transitions as a fundamental physical mechanism for charge redistribution in closely spaced multilayer systems with narrow electronic bands.

We acknowledge helpful discussions with Ady Stern and Allan MacDonald. The work at The University of Texas at Austin was supported by the Army Research Office under Grant W911NF-22-1-0160, Lyda Hill Philanthropies, the National Science Foundation (NSF) Grants MRSEC DMR-2308817 and ECCS-2122476, and the Welch Foundation Grant F-2169-20230405.
K.W. and T.T. acknowledge support from the JSPS KAKENHI (Grants 21H05233 and 23H02052), the CREST (JPMJCR24A5), and JST and World Premier International Research Center Initiative (WPI), MEXT, Japan. U.G. acknowledges financial support by the European Research Council (ERC) under grant QuantumCUSP (Grant Agreement No. 101077020).

\section*{Methods}

The graphene double layer heterostructure is assembled using a layer-by-layer dry transfer process similar to Ref. \cite{KyoungHighRotationalAccuracy2016}. We mechanically exfoliate a single large-area monolayer graphene flake, as well as several hexagonal boron-nitride (hBN) flakes to be used as gate dielectrics \cite{hBNSubstrates2010} and the tunnel barrier. The large-area graphene flake is partitioned into the top and bottom electrodes using electron beam lithography and plasma etching. The flakes are vacuum annealed at 410$^\circ$C to remove polymer residues. Atomic force microscopy (AFM) is performed to confirm the thickness of the hBN flakes. A hemispherical stamp is used to successively pick up the top gate dielectric hBN, the top graphene electrode, the hBN tunnel barrier, and the bottom graphene electrode; this stack is deposited onto a pre-assembled bottom gate stack. Because the top and bottom graphene electrodes originate from the same large-area flake, the twist angle between the graphene electrodes is controlled during the transfer process. Electron beam lithography, plasma etching, and metal evaporation are subsequently performed to define the sample geometry and to establish independent edge contacts \cite{OneDimensionalEdgeContact2013} to each graphene layer. 

Tunneling measurements were performed in a variable-temperature liquid $^4$He cryostat with a base temperature of 1.5 K. The tunneling conductance ($g_{\mathrm{Int}}$) is measured with a low-frequency lock-in technique where a small excitation at frequency 11 Hz is combined with a d.c. voltage bias and applied to the top graphene electrode, while the bottom graphene electrode is held at ground. Multiple contacts to the two graphene layers enable four-probe interlayer voltage ($V_{\mathrm{Int}}$) measurements. The tunneling current ($I_{\mathrm{Int}}$) is found by integrating the differential conductance, or by d.c. measurement using a semiconductor parameter analyzer; both techniques yield the same result.

\clearpage

\clearpage

\clearpage

\clearpage

\clearpage

\appendix

\setcounter{figure}{0}
\renewcommand{\thefigure}{S\arabic{figure}}

\section{Sample twist angle determination}

The twist angle between the graphene electrodes is found by examining resonant tunneling characteristics at zero magnetic field. A twist between the two layers’ crystal axes introduces a relative shift $\Delta\mathbf{K} = \hat{z} \times \theta \mathbf{K}$ of each layer’s $\mathbf{K}$ point, and a corresponding momentum mismatch $\hbar \Delta \mathbf{K}$ during tunneling which leads to tunneling resonances whose feature positions depend on the twist angle \cite{ManchesterTwistControlled2014, exp_tunneling_Greenaway_tbg_2015}. To quantitatively determine the twist angle, we employ a single particle perturbative tunneling model \cite{ZhengTunnelingParallelTwoDim1993, FeenstraSingleParticleTunnelingGrapheneGraphene2012, SergioDeLaBerreraGrTunneling2014, WillDoubleBilayer2017, NitinQuantumLifeSpectroscopy2021, KennethTunnelingEnergyhBNBarrier2025}. The band alignment of the top (bottom) electrode is described by an electrostatic potential $\phi_{\mathrm{T}}$ ($\phi_{\mathrm{B}}$) relative to ground, 
\begin{eqnarray}
C_{\mathrm{Int}} \Big( \phi_{\mathrm{T}} - \phi_{\mathrm{B}} \Big) - C_{\mathrm{TG}} \Big( V_{\mathrm{TG}} -\phi_{\mathrm{T}} \Big) = Q_{\mathrm{T}} \Big(  \mu_{\mathrm{T}} , \phi_{\mathrm{T}} \Big)
\end{eqnarray}
\begin{eqnarray}
C_{\mathrm{Int}} \Big(  -\phi_{\mathrm{T}} + \phi_{\mathrm{B}} \Big) - C_{\mathrm{BG}} \Big(V_{\mathrm{BG}}  -\phi_{\mathrm{B}} \Big) = Q_\mathrm{B} \Big(  \mu_{\mathrm{B}} , \phi_{\mathrm{B}} \Big)
\end{eqnarray}
where $\mu_{\mathrm{T}}$ ($\mu_{\mathrm{B}}$) is the top (bottom) layer chemical potential, $Q_{\mathrm{T}}$ ($Q_{\mathrm{B}}$) the charge per unit area on the top (bottom) electrodes, $C_{\mathrm{TG}}$ ($C_\mathrm{BG}$) the top (bottom) gate capacitance, and $C_\mathrm{Int}$ the interlayer capacitance. The tunneling current is, 
\begin{equation}
I_{\mathrm{Int}} = -e \displaystyle \int_{-\infty}^{\infty} T(E) \; \Big[ f(E - \mu_T) -f(E - \mu_B) \Big] \; dE
\end{equation}
where $E$ is the energy and $f(E)$ is the Fermi-Dirac distribution. The tunneling rate $T(E)$ is given by
\begin{equation}
T(E) = \dfrac{2\pi}{\hbar} \sum_{\mathbf{k}_{\mathrm{T}}, \mathbf{k}_{\mathrm{B}}; s, s'} |t|^2 A_{\mathrm{T}}(\mathbf{k}_{\mathrm{T}}, E) A_{\mathrm{B}}(\mathbf{k} _{\mathrm{B}},E) \; \delta_{ \mathbf{k}_{\mathrm{T}}, \mathbf{k}_{\mathrm{B}} }
\end{equation}
where $t$ is the interlayer coupling, $\mathbf{k}_{\mathrm{T}}$ ($\mathbf{k}_{\mathrm{B}}$) is the top (bottom) layer momentum, $\Delta \mathbf{K} = \mathbf{k}_{\mathrm{T}} - \mathbf{k}_{\mathrm{B}}$ the momentum mismatch, $A_{\mathrm{T,B}}(\mathbf{k}_{\mathrm{T,B}}, E)$ the Lorentzian spectral density of the top and bottom layers, 
\begin{equation}
A_{\mathrm{T,B}}(\mathbf{k}_{\mathrm{T,B}}, E) = \dfrac{1}{\pi} \dfrac{\Gamma}{\left( E - \epsilon_{\mathrm{T,B}}(\mathbf{k}_{\mathrm{T,B}}) \right)^2 +\Gamma^2}
\end{equation}
and $\epsilon_{\mathrm{T,B}}(\mathbf{k})=-e\phi_{\mathrm{T,B}}+\epsilon(\mathbf{k}_{\mathrm{T,B}})$ are the top and bottom layers energy-momentum dispersion, and $\Gamma$ is the quasiparticle state energy broadening. 

\begin{figure*}[t]
\center\includegraphics[width=\textwidth]{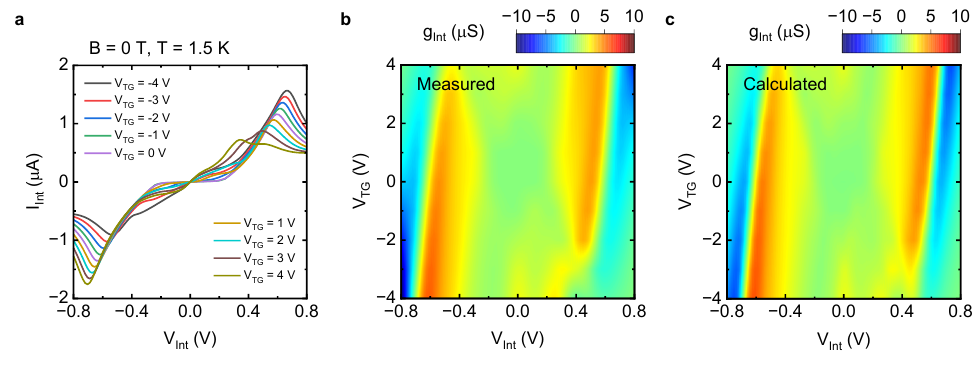}
\caption{\label{fig:S1}
\textbf{Interlayer tunneling at zero magnetic field and twist angle determination.}
\textbf{a,} Measured $I_{\mathrm{Int}}$ as a function of $V_{\mathrm{Int}}$ for various $V_{\mathrm{TG}}$ values, taken at $T = 1.5$ K and $B = 0$ T. 
\textbf{b--c,} Measured (panel \textbf{b}) and calculated (panel \textbf{c}) $g_{\mathrm{Int}} = dI_{\mathrm{Int}}/dV_{\mathrm{Int}}$ as a function of $V_{\mathrm{Int}}$ and $V_{\mathrm{TG}}$. A comparison of measured and calculated $g_{\mathrm{Int}}$ confirms the $1^\circ$ twist angle. 
}
\end{figure*}

Figure \ref{fig:S1}a shows $I_{\mathrm{Int}}$ as a function of $V_{\mathrm{Int}}$ measured at $B = 0$ T and $T = 1.5$ K, for various $V_{\mathrm{TG}}$ values with $V_{\mathrm{BG}}$ fixed at ground. Figure \ref{fig:S1}b shows $g_{\mathrm{Int}} =d I_{\mathrm{Int}} / d V_{\mathrm{Int}}$ for the same biasing conditions. Two symmetric peaks in the $g_{\mathrm{Int}}$ data evolve with $V_{\mathrm{Int}}$ and $V_{\mathrm{TG}}$; the two peaks mark the resonance condition where the two layer’s displaced Dirac cones intersect on a line, and are a tell-tale signature of tunneling with an interlayer twist angle. Figure \ref{fig:S1}c shows calculations performed to best fit the measured data of Figs. \ref{fig:S1}a--b. We find that the quasiparticle state broadening $\Gamma = 19$ meV, the interlayer hopping energy $t = 6.9 \;\mu$eV, and the twist angle $\theta = 1.0^{\circ}$ best fit the data. The excellent agreement between the measured and calculated resonant tunneling characteristics at $B = 0$ T confirms a $\theta = 1.0^{\circ}$ twist angle between the two graphene electrodes. 

\section{Charge transfer at $\nu = 0$ with nonzero charge imbalance}

\begin{figure*}
\center\includegraphics[width=0.6\textwidth]{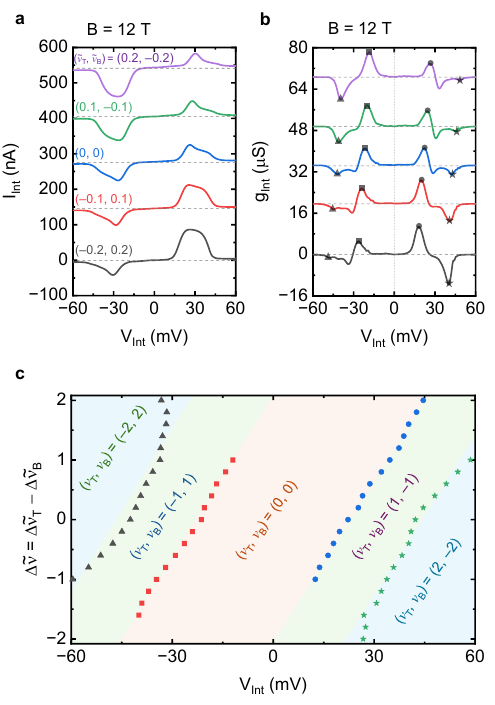}
\caption{\label{fig:S2}
\textbf{Interaction-driven charge transfer at $\nu = 0$ with charge imbalance.}
\textbf{a--b,} Measured $I_{\mathrm{Int}}$ and $g_{\mathrm{Int}}$ versus $V_{\mathrm{Int}}$ at $B = 12$ T and $T = 1.5$ K. We label the zero-bias configuration of each trace as $(\tilde{\nu}_{\mathrm{T}}, \tilde{\nu}_{\mathrm{B}})$ where $\tilde{\nu}_{\mathrm{T}} = - \tilde{\nu}_{\mathrm{B}}$. The curves are offset for clarity. 
\textbf{c,} Phase diagram mapping the top and bottom layer filling factor $(\nu_{\mathrm{T}}, \nu_{\mathrm{B}})$ versus $V_{\mathrm{Int}}$ and charge imbalance of the zero-bias configuration $\Delta \tilde{\nu} = \tilde{\nu}_{\mathrm{T}} - \tilde{\nu}_{\mathrm{B}}$. The symbols denote the center voltages of the coherent phases that separate the subband-polarized phases, and correspond to the conductance peaks marked in the panel \textbf{b} data.
}
\end{figure*}

Here we examine interaction-driven charge transfer at $\nu = 0$ for zero-bias configurations where there exists a nonzero charge imbalance, that is $\tilde{\nu}_{\mathrm{T}} = - \tilde{\nu}_{\mathrm{B}}$ at $V_{\mathrm{Int}} = 0$. Here, $\tilde{\nu}_{\mathrm{T}}$ and $\tilde{\nu}_{\mathrm{B}}$ denote the top and bottom layer filling factors at $V_{\mathrm{Int}} = 0$, respectively, and are controlled by the top and bottom gates such that $\tilde{\nu} = \tilde{\nu}_{\mathrm{T}} + \tilde{\nu}_{\mathrm{B}} = 0$. Figure \ref{fig:S2}a shows $I_{\mathrm{Int}}$ as a function of $V_{\mathrm{Int}}$ measured at $B = 12$ T and $T = 1.5$ K, for imbalanced base configurations $(\tilde{\nu}_{\mathrm{T}}, \tilde{\nu}_{\mathrm{B}}) = (0.2, -0.2)$ to $(-0.2, 0.2)$. Figure \ref{fig:S2}b shows $g_{\mathrm{Int}}$ as a function of $V_{\mathrm{Int}}$ for the same biasing conditions. We observe peaks in the $g_{\mathrm{Int}}$ versus $V_{\mathrm{Int}}$ data (symbols in Fig. \ref{fig:S2}b) that mark the onset of interaction driven charge-transfer transitions, similar to those observed in Figs. 2--3 in the main text. At near $V_{\mathrm{Int}} = 0$, the system is in the subband-polarized base configuration $(\nu_{\mathrm{T}}, \nu_{\mathrm{B}}) = (0, 0)$ where the lower two spin-valley subbands are filled and the upper two are empty (Fig. \ref{fig:S2}c, orange shaded). As $V_{\mathrm{Int}} = 0$ increases, the system enters a configuration where interlayer coherence allows a continuous charge transfer transition (Fig. \ref{fig:S2}c, circles). Beyond the interlayer coherent phase, the entire bottom layer’s second lowest spin-valley subband is transferred to the top layer, pinning the system once again to the subband-polarized configuration $(\nu_{\mathrm{T}}, \nu_{\mathrm{B}}) = (1, -1)$ [Fig. \ref{fig:S2}c, green shaded]. Figure \ref{fig:S2}c tracks the evolution of the charge transfer transitions versus charge imbalance $\Delta \tilde{\nu} = \tilde{\nu}_{\mathrm{T}} - \tilde{\nu}_{\mathrm{B}}$ at zero bias of the $\nu = 0$ state. The symbols, corresponding to the peaks marked in Fig. \ref{fig:S2}b, identify the boundaries of the subband-polarized configurations. Even at large $\Delta \tilde{\nu}$, charge imbalance of the zero-bias configuration preserves the interaction-driven charge transfer transitions, merely shifting the location of the subband-polarized and coherent phases that bridge them. 

\section{Mean-field energy, occupations, and coherence}
\label{SI_sec: occupations}

We consider a graphene double-layer separated by a tunneling barrier of thickness $d$, maintained at a fixed total filling factor $\nu$, and subjected to a perpendicular magnetic field B. We assume a sufficiently thin barrier so that Coulomb interactions remain strong while the bare interlayer tunneling can be neglected. An interlayer bias voltage $V$ is applied between the two layers. We further assume a sufficiently strong magnetic field, small filling factors, and a sufficiently weak interlayer bias so that the ground state is completely determined by the four spin-valley subbands of the $N=0$ Landau level.
The projected Hilbert space is spanned by the states $\alpha = (\ell, \tau, s)$, where $\ell=T,B$ labels the layer, while $\tau$ and $s$ denote the valley and spin degrees of freedom. The equilibrium state is determined by the competition between the single-particle Hamiltonian, the Coulomb interaction, and the applied interlayer bias.

The single-particle Hamiltonian is determined by the Zeeman splitting
$\Delta_Z=g_s\mu_BB$, the valley splitting
$\Delta_K=g_v\mu_BB$, and the applied interlayer bias $V$. The Coulomb
interaction, projected onto the $N=0$ Landau level, is
\begin{equation}
H_{\rm int}
=
\frac{1}{2A}
\sum_{\mathbf q\neq0}
\sum_{\alpha,\beta}
V_{\ell_\alpha\ell_\beta}(q)
:\rho_\alpha(-\mathbf q)\rho_\beta(\mathbf q):,
\label{Hint_density}
\end{equation}
where $A$ is the sample area, $\rho_\alpha(\mathbf q)$ is the projected density
operator for internal state $\alpha$, and
\begin{equation}
V_{\ell\ell}(q)
=
\frac{2\pi e^2}{\epsilon q},
\qquad
V_{TB}(q)
=
V_{\ell\ell}(q)e^{-qd},
\end{equation}
are the intralayer and interlayer Coulomb interactions, respectively. The
$\mathbf q=0$ component is omitted because it is cancelled by the uniform
neutralizing background, removing the long-range Hartree divergence.

Within the Hartree--Fock approximation we assume a translationally invariant one-body density matrix, which maximizes the exchange gain. The resulting mean-field energy is \cite{highB_th_Fertig_spectrum_layered_highB_1989, qhf_th_Joglekar_bias_induced_phasr_transition_2002}
\begin{equation}
\mathcal E_{\rm MF} =
\sum_{\ell,i}\epsilon_i n_{\ell i}
+\frac{v_x}{2}(\Delta\nu)^2
-\frac{\Gamma_A}{2}\sum_{\ell,i}n_{\ell i}^2
-\Gamma_E m^2
-\frac{eV}{2}\Delta\nu .
\label{Emf}
\end{equation}
The first term is the single-particle energy, where
$\epsilon_i=\mu_BB(\pm g_v\pm g_s)/2$ are the four spin-valley subband energies
within each layer. The second term is the electrostatic charging energy,
determined by the layer polarization
$\Delta\nu=\nu_T-\nu_B$ and by
\begin{equation}
v_x = \frac{d}{2\ell_B} \frac{\epsilon_\parallel}{\epsilon_\perp}
E_c,  \qquad  
E_c = \frac{e^2}{\epsilon_\parallel\ell_B},
\end{equation}
where $E_c$ is the Coulomb energy scale. The occupations
$0\le n_{\ell i}\le1$ determine  the single-particle contribution to the energy and the layer filling factors $\nu_\ell=\sum_i n_{\ell i}$, while the interlayer coherence is described by the order parameter $m$ introduced below. The applied bias drives charge transfer between the two layers.

The third and fourth terms describe the intralayer and interlayer exchange energies. Intralayer exchange favors occupation of fewer spin-valley subbands, whereas interlayer exchange favors coherent superpositions of the two layers. The corresponding interaction strengths are
\begin{equation}
\Gamma_A = \sqrt{\frac{\pi}{2}}\,E_c,
\qquad
\Gamma_E = \Gamma_A \exp\!\left(\frac{d^2\epsilon_\parallel}      {2\ell_B^2\epsilon_\perp} \right)
\operatorname{erfc}\!\left(\frac{d}{\ell_B}
\sqrt{\frac{\epsilon_\parallel}{2\epsilon_\perp}} \right),
\end{equation}
where $\ell_B$ is the magnetic length and $\epsilon_\parallel$ and $\epsilon_\perp$ are the in-plane and out-of-plane
dielectric constants of the environment. The interlayer coherence is denoted by $m$. Since the eigenvalues of the one-body density matrix satisfy $0\le\lambda\le1$, the coherence is bounded by
\begin{equation}
m^2 \le \min\!\left\{n_Tn_B,\, (1-n_T)(1-n_B) \right\},
\label{m_contraint}
\end{equation}
where $n_T$ and $n_B$ are the occupations of the two partially occupied subbands participating in the coherent superposition.

In this work, we consider a layer-balanced system. At zero bias, the two lowest
spin-valley subbands of the $N=0$ Landau level are occupied in each layer,
while the upper two subbands are empty. As the interlayer bias is increased,
charge is transferred from the bottom layer to the top layer. We denote by
$(m,n)$ the occupation sector in which subband $m$ is the highest occupied
subband in the top layer and subband $n$ is the highest occupied subband in the
bottom layer. The convention used in the main text labels the same sectors by
$(m-2,n-2)$. As the bias is increased, the ground state evolves through the
sequence
\begin{equation}
(2,2)\longrightarrow(3,2)_{\rm coh} \longrightarrow(3,1)\longrightarrow(4,1)_{\rm coh} \longrightarrow(4,0).
\end{equation}
The intermediate sectors $(3,2)$ and $(4,1)$ correspond to coherent charge-transfer states, while $(2,2)$, $(3,1)$, and $(4,0)$ are layer-polarized incoherent states.

We determine the ground state by minimizing the mean-field energy~\eqref{Emf}
within each occupation sector and then comparing the resulting minima between
sectors. To illustrate the procedure, we consider the $(3,2)$ sector. We
parameterize the occupations by $n_{T3}=x$ and $n_{B2}=1-x$, where
$0\le x\le1$, while all other occupations remain fixed at their zero-bias
values. Since the interlayer exchange contribution is negative, the energy is
minimized by the largest coherence allowed by the constraint~\eqref{m_contraint}.
We therefore set $m^2=x(1-x)$.
The mean-field energy then becomes
\begin{equation}
\mathcal E_{32}(x) =
\mathcal E_{22} +\left(eV_{32}^- - eV\right)x +\mathcal A x^2,
\label{E32}
\end{equation}
where $x$ represents the charge transferred from the
bottom layer to the top layer. Here
$\mathcal E_{22}=2\epsilon_1+2\epsilon_2-2\Gamma_A$ is the energy of the
$(2,2)$ sector,
$eV_{32}^-=\epsilon_3-\epsilon_2+\Gamma_A-\Gamma_E$, and
$\mathcal A=2v_x-\Gamma_A+\Gamma_E$. The positivity of $\mathcal A$ ensures that $\mathcal E_{32}(x)$ is convex, so its minimum occurs at
\begin{equation}
x_*=
\frac{eV-eV_{32}^-}{2\mathcal A}.
\label{xstar_32}
\end{equation}
Because the physical occupation satisfies $0\le x\le1$, the minimum is pinned to $x_*=0$ for $eV<eV_{32}^-$ and to $x_*=1$ for
$eV>eV_{32}^-+2\mathcal A$, corresponding to the incoherent $(2,2)$ and $(3,1)$ sectors, respectively. In the intermediate bias range, $eV_{32}^-<eV<eV_{32}^-+2\mathcal A$, the minimum lies in the interior of the interval, $0<x_*<1$. The system then
develops finite interlayer coherence, while the charge transfer evolves continuously with the applied bias according to Eq.~\eqref{xstar_32}.

Similar behavior occurs for the transition between the $(3,1)$ and $(4,0)$ sectors. Repeating the same analysis, we find that the two subband-polarized sectors are again separated by a coherent $(4,1)$ sector, in which the interlayer coherence is finite and the charge transfer evolves continuously with the applied bias. The occupation of the fourth top-layer subband is 
\begin{equation}
n_{T4} = \frac{eV-eV_{41}^-}{2\mathcal A},
\end{equation}
where $eV_{41}^- = \epsilon_4-\epsilon_1+4v_x+\Gamma_A-\Gamma_E$.

The midpoint voltages of the two coherent sectors, which also correspond to the peak and dip in the tunneling conductance data, are
\bea
eV_{(2,2)\to (3,1)} &=& \epsilon_3-\epsilon_2+2v_x \\
eV_{(3,1)\to (4,0)} &=& \epsilon_4-\epsilon_1+6v_x.
\label{V_midpoint}
\eea
The bias ranges corresponding to the different occupation
sectors are summarized in
Tables~\ref{tab:qp_transition_voltages}.

\begin{table}[ht]
\caption{Transition voltages for different sectors.}
\label{tab:qp_transition_voltages}

\begin{tabular}{|c|c|}
\hline
	Sector& Voltage range \\
	\hline
	$(2,2)$ & $eV\le eV_{32}^-$
	 \\[1.0ex]
	
	$(3,2)_{\rm coh}$  & $eV_{32}^- \le eV\le eV_{32}^- + 2\cA$
	 \\[1.0ex]
	
	$(3,1)$& $eV_{32}^- + 2\cA \le eV\le eV_{41}^-$
	\\[1.0ex]
	
	$(4,1)_{\rm coh}$& $eV_{41}^- \le eV\le eV_{41}^- + 2\cA$
	\\[1.0ex]

    $(4,0)$ & $eV\ge eV_{41}^- + 2\cA$\\[2.0ex]
	\hline
    & $eV_{32}^- = \epsilon_3-\epsilon_2 + \Gamma_A - \Gamma_E$ \\
    & $eV_{41}^- = \epsilon_4-\epsilon_1 + 4v_x + \Gamma_A - \Gamma_E$ \\
	& $\cA = 2v_x - \Gamma_A + \Gamma_E$ \\
    \hline
\end{tabular}
\end{table}

\section{Quasiparticle energies}
\label{SI_sec: qp energies}

The strong Coulomb interaction between the two layers implies that the quasiparticle spectrum is determined by the collective state of the graphene double-layer rather than by independent quasiparticle spectra of the individual layers.

In the subband-polarized sectors, the Hartree--Fock density matrix is diagonal
in the layer index. We separate the energy functional minimized at fixed
interlayer bias as
\begin{equation}
\mathcal E_{\rm MF} = \mathcal E_{\rm HF} -
\frac{eV}{2}\Delta\nu,
\end{equation}
where $\mathcal E_{\rm HF}$ denotes the intrinsic Hartree--Fock energy, excluding the explicit coupling to the applied bias. The quasiparticle energies, measured relative to the electrochemical potential of each layer, are obtained by differentiating $\mathcal E_{\rm HF}$ with respect to the
occupations, $\epsilon_{\ell i}^{\rm qp} = \frac{\partial\mathcal E_{\rm HF}}
{\partial n_{\ell i}}$. Using Eq.~\eqref{Emf} gives
\bea
\epsilon_{Ti}^{\rm qp} &=& \epsilon_i -\Gamma_A n_{Ti}
+ v_x\Delta\nu,
\label{qp_pol_eq_1} \\
\epsilon_{Bi}^{\rm qp} &=& \epsilon_i -\Gamma_A n_{Bi}
- v_x\Delta\nu,
\label{qp_pol_eq_2}
\eea
where $\Delta\nu=\nu_T-\nu_B$ is the layer polarization.

In the coherent sectors this procedure is no longer valid because the
Hartree--Fock energy depends explicitly on the interlayer coherence. The
appropriate variational object is the one-body density matrix,
$\rho_{\alpha\beta} = \langle c_\alpha^\dagger c_\beta \rangle$,
and the Hartree--Fock Hamiltonian is obtained from the functional derivative
\begin{equation}
h_{\alpha\beta}^{\rm HF}
=
\frac{\partial\mathcal E_{\rm HF}}
{\partial\rho_{\alpha\beta}},
\end{equation}
where all matrix elements of $\rho$ are treated as independent variables. The quasiparticle energies are therefore the eigenvalues of the Hartree--Fock Hamiltonian rather than the derivatives of the constrained mean-field energy.

As an example, consider the coherent $(3,2)$ sector. The active part of the
density matrix is
\begin{equation}
\rho_{23}
=
\begin{pmatrix}
n_{T3} & m_{23} \\
m_{23}^* & n_{B2}
\end{pmatrix},
\end{equation}
and the corresponding contribution to the Hartree--Fock energy is
\begin{equation}
\mathcal E_{23}^{\rm act} =
\epsilon_3n_{T3} + \epsilon_2n_{B2}
+ \frac{v_x}{2}\lp \Delta \nu \rp^2
- \frac{\Gamma_A}{2} \left(n_{T3}^2+n_{B2}^2\right)
- \Gamma_E|m_{23}|^2,
\end{equation}
where $\Delta \nu =2x$, and the constant contribution from the fully occupied subbands has been
omitted. Taking the functional derivative with respect to $\rho_{23}$ yields
the Hartree--Fock Hamiltonian
\begin{equation}
h_{23}^{\rm HF}
=
\begin{pmatrix}
\epsilon_3 -\Gamma_A n_{T3} + v_x\Delta \nu
& -\Gamma_E m_{23} \\
-\Gamma_E m_{23}^* &
\epsilon_2 -\Gamma_A n_{B2} - v_x\Delta \nu
\end{pmatrix}.
\end{equation}
Its eigenvalues,
\begin{equation}
E_{32,\pm}
=
\frac{\epsilon_2+\epsilon_3-\Gamma_A}{2}
\pm
\sqrt{
\left[
\frac{
\epsilon_3-\epsilon_2
+\Gamma_A
+
2(2v_x-\Gamma_A)x
}{2}
\right]^2
+
\Gamma_E^2x(1-x)
},
\end{equation}
are the quasiparticle energies that replace the uncoupled $T3$ and $B2$
branches. The remaining six quasiparticle energies are unchanged and are given
by Eqs.~\eqref{qp_pol_eq_1} and~\eqref{qp_pol_eq_2}. The quasiparticle energies
in the coherent $(4,1)$ sector are obtained in the same way.

The interlayer bias influences the quasiparticle spectrum only indirectly by changing the self-consistent occupations and coherence determined in the previous section. The quasiparticle energies of the different sectors are summarized in Tables~\ref{tab:qp_polarized}, and \ref{tab:qp_coh}.

\begin{table}[h]
\caption{Quasiparticle energies in the subband polarized sectors.}
\label{tab:qp_polarized}

\begin{tabular}{|c|c|c||c|c||c|c|}
	\hline
	\multicolumn{3}{|c||}{Sector $(2,2)$} &
    \multicolumn{2}{c||}{Sector $(3,1)$} &
    \multicolumn{2}{c|}{Sector $(4,0)$}\\
	\hline
	Subband $i$ & $T$ & $B$ & $T$ & $B$ & $T$ & $B$ \\
	\hline
	1 &
	$\epsilon_1-\Gamma_A\,$ &
	$\epsilon_1-\Gamma_A\,$ & 
    $\epsilon_1-\Gamma_A+2v_x\,$ &
	$\epsilon_1-\Gamma_A-2v_x\,$ &
	$\epsilon_1-\Gamma_A+4v_x\,$ &
	$\epsilon_1-4v_x$\\[1.0ex]
	
	2 &
	$\epsilon_2-\Gamma_A\,$ &
	$\epsilon_2-\Gamma_A\,$ &
	$\epsilon_2-\Gamma_A+2v_x\,$ &
	$\epsilon_2-2v_x\,$&
	$\epsilon_2-\Gamma_A+4v_x\,$ &
	$\epsilon_2-4v_x$ \\[1.0ex]
	
	3 &
	$\epsilon_3\,$ &
	$\epsilon_3\,$ &
	$\epsilon_3-\Gamma_A+2v_x\,$ &
	$\epsilon_3-2v_x\,$&
	$\epsilon_3-\Gamma_A+4v_x\,$ &
	$\epsilon_3-4v_x$\\[1.0ex]
	
	4 &
	$\epsilon_4\,$ &
	$\epsilon_4\,$ &
	$\epsilon_4+2v_x\,$ &
	$\epsilon_4-2v_x\,$&
	$\epsilon_4-\Gamma_A+4v_x\,$ &
	$\epsilon_4-4v_x$\\[2.0ex]
	\hline
\end{tabular}
\end{table}
\begin{table}
\centering
\caption{Quasiparticle energies in the coherent sectors.}
\label{tab:qp_coh}

\begin{tabular}{|c|c|c||c|c|}
	\hline  
    \multicolumn{1}{|c|}{ }&
	\multicolumn{2}{c||}{Sector $(3,2)_{\rm coh}$}&
	\multicolumn{2}{c|}{Sector $(4,1)_{\rm coh}$} \\
	\hline
	$\textrm{Subband}\,i$& $T$ & $B$& $T$ & $B$ \\
	\hline
	1 &
	$\epsilon_1-\Gamma_A+2v_xx$ &
	$\epsilon_1-\Gamma_A-2v_xx$&
	$\epsilon_1-\Gamma_A+2v_x(1+y)$ &
	 \\[1.0ex]
	
	2 &
	$\epsilon_2-\Gamma_A+2v_xx$ &
	&
	$\epsilon_2-\Gamma_A+2v_x(1+y)$ &
	$\epsilon_2-2v_x(1+y)$ \\[1.0ex]
	
	3 &
	 &
	$\epsilon_3-2v_xx$&
	$\epsilon_3-\Gamma_A+2v_x(1+y)$ &
	$\epsilon_3-2v_x(1+y)$\\[1.0ex]
	
	4&
	$\epsilon_4+2v_xx$ &
	$\epsilon_4-2v_xx$&
	 &
	$\epsilon_4-2v_x(1+y)$\\[2.0ex]
	\hline
	\multicolumn{1}{|c|}{}&
    \multicolumn{2}{c||}{$\makecell{E_{32,\pm}
=
\frac{\epsilon_2+\epsilon_3-\Gamma_A}{2}
\pm\\
\sqrt{
\left[
\frac{
\epsilon_3-\epsilon_2
+\Gamma_A
+
2(2v_x-\Gamma_A)x
}{2}
\right]^2
+
\Gamma_E^2x(1-x)
}}$}&
    \multicolumn{2}{c|}{$\makecell{E_{41,\pm}
=
\frac{\epsilon_1+\epsilon_4-\Gamma_A}{2}
\pm\\
\sqrt{
\left[
\frac{
\epsilon_4-\epsilon_1
+\Gamma_A
+4v_x
+
2(2v_x-\Gamma_A)y
}{2}
\right]^2
+
\Gamma_E^2y(1-y)
}}$} \\
	\hline
	\multicolumn{1}{|c|}{}&
    \multicolumn{2}{c||}{$x
	=
	\frac{
		eV-(\epsilon_3-\epsilon_2+\Gamma_A-\Gamma_E)
	}{
		2(2v_x-\Gamma_A+\Gamma_E)
	}$}&
    \multicolumn{2}{c|}{$y
	=
	\frac{
		eV-(\epsilon_4-\epsilon_1+4v_x+\Gamma_A-\Gamma_E)
	}{
		2(2v_x-\Gamma_A+\Gamma_E)
	}$} \\
	\hline
\end{tabular}
\end{table}

\section{Projected tunneling Hamiltonian}
\label{SI_sec: projected tunneling hamiltonian}

The continuum tunneling Hamiltonian of twisted double-layer graphene is
\begin{equation}
H_{\rm tunnel} = \int d^2r\; \psi_T^\dagger(\mathbf r)\, T(\mathbf r)\, \psi_B(\mathbf r)
+ {\rm h.c.},
\label{eq:HT_continuum}
\end{equation}
where $\psi_{\ell}(\bm{r})$ is the electron annihilation operator at position $\bm{r}$ in layer $\ell=T,B$, and $T(\mathbf r) = \sum_{j=1}^{3} t_j e^{-i\mathbf q_j\cdot\mathbf r}$ \cite{moireBands}.
Here \(t_j\) are the three moiré tunneling matrices and
\(\mathbf q_j\) are the corresponding momentum-transfer vectors,
$\mathbf q_1 = k_\theta(0,-1)$,
$\mathbf q_2 = \frac{k_\theta}{2}(\sqrt3,1)$,
$\mathbf q_3 = \frac{k_\theta}{2}(-\sqrt3,1)$.
For small angles $k_\theta \simeq K_D\theta$ with $K_D = \frac{4\pi}{3a}$ the Dirac momentum and $a$ the graphene lattice constant. Throughout this work we assume that tunneling preserves both spin and valley.
We therefore derive the tunneling Hamiltonian for a single spin--valley flavor;
the total current is obtained by summing over all four flavors.

We project Eq.~\eqref{eq:HT_continuum} onto the $N=0$ Landau-level states in the two layers. In the Landau gauge $\mathbf A = B(0,x)$, the guiding-center basis states are labeled by $X = -k_y\ell_B^2$, with $\ell_B = \sqrt{\frac{\hbar}{eB}}$ the magnetic length.
For a fixed valley $\tau$, the graphene \(N=0\) Landau level resides on a single sublattice. The sublattice structure of the tunneling matrices therefore reduces to valley-dependent phase factors and the projected field operators
take the form $\psi_\ell(\mathbf r) = \sum_X \phi_{0X}(\mathbf r)\,c_{\ell X}$ where $\phi_{0X}$ is the lowest Landau-level wavefunction. Using 
\begin{equation}
\langle 0,X_T|e^{-i\mathbf q\cdot\mathbf r} |0,X_B\rangle
= e^{-q^2\ell_B^2/4}
e^{-iq_x(X_1+X_2)/2}
\delta_{X_T,X_B-q_y\ell_B^2},
\label{eq:guiding_center_shift}
\end{equation}
we find that tunneling by momentum \(\mathbf q_j\) shifts the guiding center by
$q_{j,y}\ell_B^2$ so that the projected tunneling Hamiltonian is \cite{moireButterflies}
\begin{equation}
H_{\rm tunnel}^{(0)} =  \sum_{j=1}^{3} \sum_X T_{jX}
c^\dagger_{2,X-q_{j,y}\ell_B^2} c_{1,X}
+ {\rm h.c.},
\label{eq:HT_projected}
\end{equation}
where
\begin{equation}
T_{jX} =\Lambda e^{-iq_{j,x}(X-q_{j,y}\ell_B^2/2)} e^{-i\tau\varphi_j}, \qquad
\Lambda = w e^{-k_\theta^2\ell_B^2/4},
\label{TjX_fixed}
\end{equation}
with $\varphi_1=0$, $\varphi_2=\frac{2\pi}{3}$, $\varphi_3=-\frac{2\pi}{3}$ and $w$ the tunneling energy.
The exponential factor is the $N=0$ Landau-level form factor associated with the momentum transfer $\mathbf q_j$. It strongly suppresses tunneling when $k_\theta\ell_B\gg1$, as occurs at large twist angles or strong magnetic fields.

\section{Tunneling current}
\label{SI_sec: tunneling current}

The tunneling current between two weakly coupled electronic systems is typically
described by a convolution of the spectral functions of the two subsystems.
In that picture, interactions between the two subsystems are
negligible, so the two spectral functions can be treated as properties of
independent layers.

This description is not applicable to double-layer quantum Hall
ferromagnets, where the interlayer Coulomb interaction is comparable to, or
larger than, the relevant single-particle energy scales. Consequently, the
occupations of the two layers cannot be determined independently. Instead, as described in Sec.~\ref{SI_sec: occupations}, for each applied bias voltage \(V\), the occupations \(n_{\ell i}(V)\) are
obtained by minimizing the interacting bilayer energy. Because the
quasiparticle energies depend on these occupations through the Hartree and exchange interactions (see Sec.~\ref{SI_sec: qp energies}), the resulting spectrum is a property of the interacting
bilayer rather than of the individual layers.

For a given bias voltage, the mean-field energy is
\begin{equation}
\mathcal E_{\rm MF} = \mathcal E_{\rm HF} - \frac{eV}{2} \left(\nu_T-\nu_B\right)
-\bar{\mu}\left( \nu_T+\nu_B-\nu \right),
\label{E_MF}
\end{equation}
where \(\mathcal E_{\rm HF}\) contains the Hartree and exchange interactions discussed in the main text. Because the interlayer capacitance is much larger than the capacitance to the gates, the total filling factor \(\nu\) is approximately independent of the interlayer bias. This constraint is enforced by the Lagrange multiplier \(\bar{\mu}\). 

Minimizing Eq.~\eqref{E_MF} determines the occupations
\(n_{\ell i}(V)\). In the gauge used below, the quasiparticle energies entering the spectral functions are defined from the internal HF functional,
\begin{equation}
\epsilon_{\ell i}^{\rm qp}(V) = \left. \frac{\partial \mathcal E_{\rm HF}} {\partial n_{\ell i}} \right|_{n=n(V)} .
\label{qp_definition}
\end{equation}
The explicit bias term in Eq.~\eqref{E_MF} is instead carried by the time-dependent phase of the tunneling operator.
After the time-dependent gauge transformation, the projected tunneling
Hamiltonian acquires the form
\begin{equation}
H_T(t) = \sum_{j,X,i} \left[ T_{jX} e^{-ieVt/\hbar} c^\dagger_{B i,X_j}(t) c_{T i,X}(t) + {\rm h.c.} \right],
\label{HT_time}
\end{equation}
where \(X_j=X-q_{j,y}\ell_B^2\). The remaining time dependence of the
operators is generated by the tunneling-free HF Hamiltonian.
The current operator from the top layer to the bottom layer is
\begin{equation}
\hat I = -e\dot N_T =
\frac{ie}{\hbar} \left[N_T,H_{T,I}(t)\right].
\label{Iop}
\end{equation}
Here the subscript \(I\) denotes the interaction picture with respect to the tunneling-free Hamiltonian.

\subsection{Exact second-order tunneling current}

The interlayer tunneling amplitude is several orders of magnitude smaller
than both the single-particle energy scales and the Coulomb interaction
energies. We therefore treat tunneling perturbatively to leading
nonvanishing order in the tunneling amplitude \(w\). The equilibrium state
of the double layer is determined entirely by the interacting tunneling-free Hamiltonian, while the tunneling Hamiltonian is treated as a perturbation.

Since the expectation value of the current vanishes in the absence of
tunneling, the leading contribution is second order in \(w\). Linear
response therefore gives
\begin{equation}
I(V) = -\frac{i}{\hbar} \int_{-\infty}^{t} dt' \,
\left< \left[ \hat I_I(t), H_{T,I}(t') \right] \right>_0,
\label{Kubo_current}
\end{equation}
where \(\langle\cdots\rangle_0\) denotes an expectation value with respect to the equilibrium state of the interacting bilayer with the weak
tunneling Hamiltonian omitted.

Substituting Eqs.~(\ref{HT_time}, \ref{TjX_fixed}) into Eq.~(\ref{Kubo_current}) gives
\be
I(V) = \frac{e w^2}{\hbar^2} e^{-k_\theta^2 \ell_B^2/2} \sum_{i i'} \sum_{j,j'=1}^{3} \int_{-\infty}^{\infty} dt \,
\left[ e^{-ieVt/\hbar} \mathcal C^{>}_{ii';jj'}(t)
- e^{ieVt/\hbar} \mathcal C^{<}_{ii';jj'}(t) \right],
\label{I_exact_jjprime}
\ee
where
\begin{equation}
\mathcal C^{>}_{ii';jj'}(t)
=
\sum_{X,X'}
e^{i\phi_{ji}(X)-i\phi_{j'i'}(X')}
\left<
c_{TiX}^{\dagger}(t)
c_{Bi,X+q_{jy}\ell_B^2}(t)
c_{Bi',X'+q_{j'y}\ell_B^2}^{\dagger}(0)
c_{Ti'X'}(0)
\right>_0 ,
\end{equation}
and
\begin{equation}
\mathcal C^{<}_{ii';jj'}(t)
=
\sum_{X,X'}
e^{-i\phi_{ji}(X)+i\phi_{j'i'}(X')}
\left<
c_{Bi,X+q_{jy}\ell_B^2}^{\dagger}(t)
c_{TiX}(t)
c_{Ti'X'}^{\dagger}(0)
c_{Bi',X'+q_{j'y}\ell_B^2}(0)
\right>_0,
\end{equation}
with  $\phi_{ji}(X) = q_{j,x}(X-q_{j,y}\ell_B^2/2) + \tau_i \varphi_j$.
It is useful to introduce the matrix of tunneling correlation functions
\begin{equation}
S^{\gtrless}_{ii';jj'}(\omega) = \int_{-\infty}^{\infty} dt\, e^{i\omega t} \mathcal C^{\gtrless}_{ii';jj'}(t).
\end{equation}
The current can then be written as
\begin{equation}
I(V) =
\frac{e w^2}{\hbar^2} e^{-k_\theta^2 \ell_B^2/2}
\sum_{i i'} \sum_{j,j'=1}^{3}
\left[ S^{>}_{ii';jj'}(-\omega_V)
- S^{<}_{ii';jj'}(\omega_V) \right],
\label{I_exact_twisted_final}
\end{equation}
where $\omega_V=eV/\hbar$.
Equation~\eqref{I_exact_twisted_final} is exact to second order in
the tunneling amplitude.

\subsection{Hartree--Fock approximation}

To obtain an explicit expression for the tunneling current, the correlation functions \(S_{ii';jj'}^{>(<)}\) appearing in  Eq.~(\ref{I_exact_twisted_final}) are evaluated within the Hartree--Fock approximation. For each bias voltage \(V\), the equilibrium state is obtained by minimizing $\mathcal E_{\rm MF}$, the Hartree--Fock energy of the interacting bilayer given by Eq.~\eqref{E_MF}.
The weak tunneling Hamiltonian is omitted from this minimization and is treated only perturbatively through Eq.~(\ref{I_exact_twisted_final}).

The self-consistent Hartree--Fock solution determines the mean-field Hamiltonian, the occupation numbers, and, where allowed, the interlayer coherence order parameters.

Within the Hartree--Fock approximation, the four-fermion correlation functions entering Eq.~(\ref{I_exact_twisted_final}) are evaluated using Wick's theorem and expressed in terms of single-particle Green's functions of the self-consistent Hartree--Fock Hamiltonian. The resulting tunneling current depends on the structure of the 
equilibrium state.

In the subband-polarized sectors, the Hartree--Fock Hamiltonian is diagonal in both layer and spin--valley indices, and the tunneling current reduces to the familiar bubble diagram constructed from layer-resolved single-particle Green's functions.

In the coherent sectors, the Hartree--Fock Hamiltonian contains off-diagonal matrix elements between the coherent subbands, and the single-particle Green's functions become matrices in the corresponding
two-level subspace. The tunneling current is again obtained from the Hartree--Fock bubble. As shown below, for the coherent states considered
here the spin--valley structure of the order parameter together with spin--valley-conserving tunneling causes the $i\neq i'$ contributions
to vanish. Interlayer coherence therefore enters the tunneling current through the reconstructed diagonal spectral functions.

\subsection{Subband-polarized sectors}

We first consider the subband-polarized sectors, in which the Hartree--Fock Hamiltonian is diagonal in the layer and spin--valley indices. Consequently, the intralayer Green's functions are diagonal in the spin--valley index, while the interlayer contractions vanish. Wick's theorem therefore gives
\be
\left< c^\dagger_{Ti}(t) c_{Bi}(t) c^\dagger_{Bi'}(0) c_{Ti'}(0) \right>
= \delta_{ii'}\left< c^\dagger_{Ti}(t) c_{Ti}(0) \right>
\left< c_{Bi}(t) c^\dagger_{Bi}(0) \right>.
\ee
Introducing the lesser and greater Green's functions,
\bea
G_{\ell i}^{<}(X,X';t) &=& i \left< c^\dagger_{\ell i,X'}(0) c_{\ell i,X}(t) \right>,   \\
G_{\ell i}^{>}(X,X';t) &=& -i \left< c_{\ell i,X}(t) c^\dagger_{\ell i,X'}(0) \right>,
\eea
Eq.~(\ref{I_exact_jjprime}) becomes
\bea
I(V) &=& \frac{e w^2}{\hbar^2} e^{-k_\theta^2\ell_B^2/2}
\sum_{i,j,j'} \int dt e^{-ieVt/\hbar} 
\sum_{XX'} e^{i\phi_j(X)-i\phi_{j'}(X')}
G^{<}_{Ti}(X',X;t) G^{>}_{Bi}(X_j,X'_{j'};-t)  \nonumber \\
&-& (T\leftrightarrow B),
\label{I_GF}
\eea
where \(X_j=X+q_{j,y}\ell_B^2\). After disorder averaging, translational invariance is restored,
\begin{equation}
\overline{ G^{\gtrless}_{\ell i}(X,X';\omega) } =
\delta_{XX'} G^{\gtrless}_{\ell i}(\omega),
\end{equation}
and the \(j\neq j'\) terms vanish. Using
\bea
G_{\ell i}^{<}(\omega) &=& i \Theta(\bar\mu-\omega) A_{\ell i}(\omega), \\
G_{\ell i}^{>}(\omega) &=& -i[1-\Theta(\bar\mu-\omega)]A_{\ell i}(\omega),
\eea
where \(\bar\mu\) is the Lagrange multiplier enforcing the fixed total
filling, the expression for the current for $eV>0$ gives
\begin{equation}
I(V) = C\sum_i \int_{\bar\mu-eV}^{\bar\mu} \frac{d\omega}{2\pi}\,
A_{Ti}(\omega) A_{Bi}(\omega+eV),
\label{I_zero_temperature}
\end{equation}
where $C= \frac{3eN_\phi}{\hbar} w^2 \exp\lp-k_\theta^2\ell_B^2/2\rp$.
We model disorder broadening phenomenologically by the Lorentzian spectral function
\begin{equation}
A_{\ell i}(\omega)
=
\frac{2\Gamma}
{(\omega-\epsilon_{\ell i}^{\rm qp})^2+\Gamma^2}
\end{equation}
with $\Gamma$ chosen to fit the experimental data. The integral in \eqref{I_zero_temperature} can be evaluated analytically, yielding 
\be
I(V) = \mathcal C \sum_i \left[ F(x_2,\Delta_i) - F(x_1,\Delta_i)
\right],
\label{I_T0_analytic}
\ee
where $x_1 = \bar\mu - eV - \epsilon_{T i}^{\rm qp}(V)$, 
$x_2 = \bar\mu - \epsilon_{T i}^{\rm qp}(V)$,
$\Delta_i(V) = \epsilon_{B i}^{\rm qp}(V) - \epsilon_{T i}^{\rm qp}(V) - eV$, and
\begin{equation}
F(x,\Delta) = \frac{2\Gamma}{\pi\left(4\Gamma^2+\Delta^2\right)}
\left[ \tan^{-1} \left( \frac{x}{\Gamma} \right) + 
\tan^{-1} \left( \frac{x-\Delta}{\Gamma} \right) \right]
+ \frac{2\Gamma^2}{\pi\Delta\left(4\Gamma^2+\Delta^2\right)}
\ln \left[ \frac{x^2+\Gamma^2} { \left( x-\Delta \right)^2+\Gamma^2 }
\right].
\label{Ffunction}
\end{equation}

\subsection{Coherent sectors}

We now consider the coherent sectors, in which interlayer exchange hybridizes subbands belonging to opposite layers and carrying different spin--valley indices. A special feature of the layer-balanced state considered here is that the interlayer coherence connects different spin--valley subbands, so that
$\langle c^\dagger_{Ta}c_{Bb}\rangle$ can be finite only if $a\neq b$. The projected tunneling Hamiltonian, by contrast, conserves the spin--valley index, so that $\langle c^\dagger_{Ti}c_{Bi}\rangle=0$ also in the coherent sectors. Moreover, the intralayer Green's functions remain diagonal in the spin--valley index. Wick's theorem therefore retains the form derived in Eq.~(\ref{I_zero_temperature}).

Interlayer coherence nevertheless reconstructs the diagonal single-particle Green's functions. The spectral function of a subband participating in a coherent pair acquires two quasiparticle poles with
coherence-dependent spectral weights. The tunneling current in the coherent sectors is therefore obtained from Eq.~(\ref{I_zero_temperature}) using these reconstructed diagonal spectral functions.

For a coherent pair of subbands, the Hartree--Fock Hamiltonian is
\begin{equation}
H_{\rm HF}
=
\begin{pmatrix}
\epsilon_T^{\rm qp} & \Delta \\
\Delta^* & \epsilon_B^{\rm qp}
\end{pmatrix},
\end{equation}
where
\(\Delta=\Gamma_E m\)
is the self-consistent interlayer coherence field.
Its eigenvalues are $E_{\pm} = \bar\epsilon \pm R$ with $R = \sqrt{\eta^2+|\Delta|^2}$, $\bar\epsilon = (\epsilon_T^{\rm qp}+\epsilon_B^{\rm qp})/2$, and $\eta = (\epsilon_T^{\rm qp}-\epsilon_B^{\rm qp})/2$. The retarded Green's function is
\begin{equation}
G^R(\omega) = (\omega-H_{\rm HF}+i\Gamma)^{-1}.
\end{equation}
Using the spectral decomposition of the Hartree--Fock Hamiltonian,
\begin{equation}
G^R(\omega)
=
\sum_{\alpha=\pm}
\frac{|u_\alpha\rangle\langle u_\alpha|}
{\omega-E_\alpha+i\Gamma},
\end{equation}
the spectral-function matrix becomes
\begin{equation}
A(\omega)
=
-2\,{\rm Im}\,G^R(\omega)
=
\sum_{\alpha=\pm}
|u_\alpha\rangle
\langle u_\alpha|
\frac{2\Gamma}
{(\omega-E_\alpha)^2+\Gamma^2}.
\end{equation}

The tunneling current depends only on the diagonal matrix elements of the spectral function,
\bea
A_T(\omega) &=& Z_+^{T} L_\Gamma(\omega-E_+)
+ Z_-^{T} L_\Gamma(\omega-E_-), \\
A_B(\omega) &=& Z_+^{B} L_\Gamma(\omega-E_+)
+ Z_-^{B} L_\Gamma(\omega-E_-),
\eea
where $L_\Gamma(x) = 2\Gamma/(x^2+\Gamma^2)$, and
\begin{align}
Z_\pm^T
&=
|\langle T|u_\pm\rangle|^2
=
\frac12\left(1\pm\frac{\eta}{R}\right),
\\
Z_\pm^B
&=
|\langle B|u_\pm\rangle|^2
=
\frac12\left(1\mp\frac{\eta}{R}\right).
\end{align}

Substituting the coherent spectral functions into
Eq.~(\ref{I_zero_temperature}) gives the tunneling current in the
coherent sectors, where \(A_{Ti}\) and \(A_{Bi}\) are the coherent spectral functions given above. The only difference from the subband-polarized sectors is therefore the replacement of the single Lorentzian quasiparticle peaks by two hybridized quasiparticle branches whose energies and spectral weights are determined self-consistently by the interlayer coherence.

\section{Charge-transfer transition width}
\label{SI_sec:transition_width}

Charge transfer occurs through an interlayer-coherent phase in which the
occupied state is a coherent superposition of the two layers. Across this
phase, the charge associated with an entire spin--valley subband of the
$N=0$ LL is continuously transferred between the layers. The corresponding
energy width of the coherent phase is
\begin{equation}
\Delta_{\rm coh}(d,B) \equiv e\Delta V_{\rm coh}
=2\mathcal{A} =2\left(2v_x-\Gamma_A+\Gamma_E\right),
\label{eq:coh_width}
\end{equation}
where $\Delta V_{\rm coh}$ is its extent in interlayer bias voltage. This
coherent charge-transfer process is illustrated in Fig.~2c(viii) of the
main text. Definitions of $v_x$, $\Gamma_A$, and $\Gamma_E$ are given in
Sec.~\ref{SI_sec: occupations}.

The transition width also provides a measure of the energetic stiffness
against charge transfer. Within the coherent phase, the layer polarization
varies linearly with bias and satisfies
\begin{equation}
\frac{d(eV_{\rm Int})}{d\Delta\nu} = \mathcal{A} = 
\frac{\Delta_{\rm coh}}{2}.
\label{eq:inverse_capacitance}
\end{equation}
Thus, $\Delta_{\rm coh}/2$ measures the change in bias
required to transfer charge between the layers and is proportional to the
inverse differential interlayer capacitance.

\begin{figure*}
\center\includegraphics[width=0.6\textwidth]{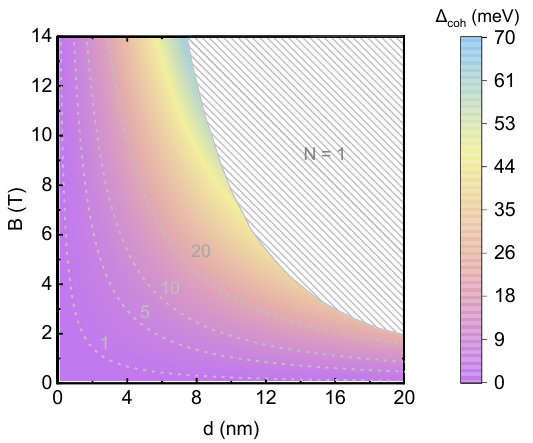}
\caption{\textbf{Effect of interlayer separation on charge transfer.}
Calculated coherent-phase width $\Delta_{\rm coh}(d,B)=e\Delta V_{\rm coh}$ as a function of interlayer separation $d$ and magnetic field $B$.
$\Delta_{\rm coh}$ gives the energy range in interlayer bias over which the interlayer-coherent phase connects neighboring subband-polarized quantum Hall ferromagnetic configurations. The dashed contours indicate
$\Delta_{\rm coh}=1$, $5$, $10$, and $20$~meV. Decreasing the interlayer
separation narrows the coherent phase and produces a sharper, more
step-like $\Delta\nu(V_{\rm Int})$ dependence. The hatched region indicates the parameter range where projection onto the $N=0$ LL is no longer accurate.}
\label{fig:S3}
\end{figure*}

Figure~\ref{fig:S3} shows $\Delta_{\rm coh}(d,B)$ over
experimentally relevant magnetic fields and interlayer separations. The
transition becomes sharper with decreasing $d$ and broader with increasing
$B$. Resolving the individual spin--valley subbands, however, requires a
sufficiently strong magnetic field. This competition is unfavorable in
conventional GaAs/AlGaAs bilayers, where typical interlayer separations are $d\sim10$--$20$~nm. In the graphene double layer studied here, by contrast, the atomic-scale separation $d=1.8$~nm maintains a narrow coherent window even at magnetic fields as large as $14$~T. Consequently, although charge transfer through the coherent phase is continuous, the narrow width of this phase produces a nearly step-like $\Delta\nu(V_{\rm Int})$ dependence.

\bibliography{references_arXiv.bib}

@incollection{review_Girvin_MacDonald_multicomponent_QH_1995,
  author = {{Girvin}, S. M. and {MacDonald}, A. H.},
  title = {Multi-Component Quantum Hall Systems: The Sum of their Parts and More},
  editor      = "Sankar Das Sarma and Aron Pinczuk",
  booktitle   = "Perspectives in Quantum Hall Effects: Novel Quantum Liquids in Low-Dimensional Semiconductor Structures",
  publisher   = "Wiley",
  year        = 1996,
  chapter     = 5,
            eid = {cond-mat/9505087},
        pages = {cond-mat/9505087},
          doi = {10.48550/arXiv.cond-mat/9505087},
archivePrefix = {arXiv},
       eprint = {cond-mat/9505087},
 primaryClass = {cond-mat},
 url={https://ui.adsabs.harvard.edu/abs/1996pqhe.book.....D}
}

@article{qhf_exp_Tutuc_density_imbalance_WC_2003,
  title = {Role of Density Imbalance in an Interacting Bilayer Hole System},
  author = {Tutuc, E. and Melinte, S. and De Poortere, E. P. and Pillarisetty, R. and Shayegan, M.},
  journal = {Phys. Rev. Lett.},
  volume = {91},
  issue = {7},
  pages = {076802},
  numpages = {4},
  year = {2003},
  month = {Aug},
  publisher = {American Physical Society},
  doi = {10.1103/PhysRevLett.91.076802},
  url = {https://link.aps.org/doi/10.1103/PhysRevLett.91.076802}, 
  archivePrefix = {arXiv},
  eprint = {cond-mat/0209649},
  primaryClass = {cond-mat.mes-hall}
}

@article{qhf_exp_Champagne_charge_imbalance_2008,
  title = {Charge imbalance and bilayer two-dimensional electron systems at ${\ensuremath{\nu}}_{T}=1$},
  author = {Champagne, A. R. and Finck, A. D. K. and Eisenstein, J. P. and Pfeiffer, L. N. and West, K. W.},
  journal = {Phys. Rev. B},
  volume = {78},
  issue = {20},
  pages = {205310},
  numpages = {12},
  year = {2008},
  month = {Nov},
  publisher = {American Physical Society},
  doi = {10.1103/PhysRevB.78.205310},
  url = {https://link.aps.org/doi/10.1103/PhysRevB.78.205310}, 
  archivePrefix = {arXiv},
  print = {0808.1257},
  primaryClass = {cond-mat.mes-hall}
}

@article{qhf_exp_Spielman_density_imbalance_2004,
  title = {Onset of interlayer phase coherence in a bilayer two-dimensional electron system: Effect of layer density imbalance},
  author = {Spielman, I. B. and Kellogg, M. and Eisenstein, J. P. and Pfeiffer, L. N. and West, K. W.},
  journal = {Phys. Rev. B},
  volume = {70},
  issue = {8},
  pages = {081303(R)},
  numpages = {4},
  year = {2004},
  month = {Aug},
  publisher = {American Physical Society},
  doi = {10.1103/PhysRevB.70.081303},
  url = {https://link.aps.org/doi/10.1103/PhysRevB.70.081303}, 
  archivePrefix = {arXiv},
  eprint = {cond-mat/0406067},
  primaryClass = {cond-mat.mes-hall}
}

@article{highB_th_GMP_1986,
  title = {Magneto-roton theory of collective excitations in the fractional quantum Hall effect},
  author = {Girvin, S. M. and MacDonald, A. H. and Platzman, P. M.},
  journal = {Phys. Rev. B},
  volume = {33},
  issue = {4},
  pages = {2481--2494},
  numpages = {0},
  year = {1986},
  month = {Feb},
  publisher = {American Physical Society},
  doi = {10.1103/PhysRevB.33.2481},
  url = {https://link.aps.org/doi/10.1103/PhysRevB.33.2481}
}

@article{highB_th_Fertig_spectrum_layered_highB_1989,
  title = {Energy spectrum of a layered system in a strong magnetic field},
  author = {Fertig, H. A.},
  journal = {Phys. Rev. B},
  volume = {40},
  issue = {2},
  pages = {1087--1095},
  numpages = {0},
  year = {1989},
  month = {Jul},
  publisher = {American Physical Society},
  doi = {10.1103/PhysRevB.40.1087},
  url = {https://link.aps.org/doi/10.1103/PhysRevB.40.1087}
}

@article{qhf_th_Joglekar_bias_induced_phasr_transition_2002,
  title = {Bias-voltage-induced phase transition in bilayer quantum Hall ferromagnets},
  author = {Joglekar, Yogesh N. and MacDonald, Allan H.},
  journal = {Phys. Rev. B},
  volume = {65},
  issue = {23},
  pages = {235319},
  numpages = {9},
  year = {2002},
  month = {Jun},
  publisher = {American Physical Society},
  doi = {10.1103/PhysRevB.65.235319},
  url = {https://link.aps.org/doi/10.1103/PhysRevB.65.235319}, 
  archivePrefix = {arXiv},
  eprint = {cond-mat/0111056},
  primaryClass = {cond-mat.mes-hall}
}

@Article{exp_Young_gs_gv_2012,
  author={Young, A. F.
  and Dean, C. R.
  and Wang, L.
  and Ren, H.
  and Cadden-Zimansky, P.
  and Watanabe, K.
  and Taniguchi, T.
  and Hone, J.
  and Shepard, K. L.
  and Kim, P.},
  title={Spin and valley quantum Hall ferromagnetism in graphene},
  journal={Nature Phys},
  year={2012},
  month={Jul},
  day={01},
  volume={8},
  number={7},
  pages={550-556},
  URL={https://doi.org/10.1038/nphys2307}, 
  doi={10.1038/nphys2307}, 
  archivePrefix = {arXiv},
  eprint = {1201.4167},
  primaryClass = {cond-mat.mes-hall},
}

@article{exp_Jiang_graphene_gs_gv_2019,
       author = {{Jiang}, Yuxuan and {Lu}, Zhengguang and {Gigliotti}, Jamey and {Rustagi}, Avinash and {Chen}, Li and {Berger}, Claire and {de Heer}, Walt and {Stanton}, Christopher J. and {Smirnov}, Dmitry and {Jiang}, Zhigang},
        title = "{Valley and Zeeman Splittings in Multilayer Epitaxial Graphene Revealed by Circular Polarization Resolved Magneto-infrared Spectroscopy}",
      journal = {Nano Letters},
         year = 2019,
        month = oct,
       volume = {19},
       number = {10},
        pages = {7043-7049},
          doi = {10.1021/acs.nanolett.9b02505},
archivePrefix = {arXiv},
       eprint = {1909.01501},
 primaryClass = {cond-mat.mes-hall},
       url = {https://doi.org/10.1021/acs.nanolett.9b02505}
}

@ARTICLE{exp_tunneling_Greenaway_tbg_2015,
       author = {{Greenaway}, M.~T. and {Vdovin}, E.~E. and {Mishchenko}, A. and {Makarovsky}, O. and {Patan{\`e}}, A. and {Wallbank}, J.~R. and {Cao}, Y. and {Kretinin}, A.~V. and {Zhu}, M.~J. and {Morozov}, S.~V. and {Fal'ko}, V.~I. and {Novoselov}, K.~S. and {Geim}, A.~K. and {Fromhold}, T.~M. and {Eaves}, L.},
        title = "{Resonant tunnelling between the chiral Landau states of twisted graphene lattices}",
      journal = {Nature Physics},
         year = 2015,
        month = dec,
       volume = {11},
       number = {12},
        pages = {1057-1062},
          doi = {10.1038/nphys3507},
archivePrefix = {arXiv},
       eprint = {1509.06208},
 primaryClass = {cond-mat.mes-hall},
       url = {https://doi.org/10.1038/nphys3507}
}

@Article{KennethTunnelingEnergyhBNBarrier2025,
author={Lin, Kenneth A.
and Ramey, Brooke
and Prasad, Nitin
and Burg, G. William
and Watanabe, Kenji
and Taniguchi, Takashi
and Register, Leonard F.
and Tutuc, Emanuel},
title={Tunneling Energy and Capacitance of Twist-Controlled Graphene Double Layers Separated by Boron Nitride Barriers},
journal={Nano Letters},
year={2025},
month={Dec},
day={10},
publisher={American Chemical Society},
volume={25},
number={49},
pages={17010-17015},
issn={1530-6984},
doi={10.1021/acs.nanolett.5c03773},
url={https://doi.org/10.1021/acs.nanolett.5c03773}
}

@ARTICLE{EmergenceOfInterlayerCoherenceInKenneth2022,
       author = {{Lin}, Kenneth A. and {Prasad}, Nitin and {Burg}, G. William and {Zou}, Bo and {Ueno}, Keiji and {Watanabe}, Kenji and {Taniguchi}, Takashi and {MacDonald}, Allan H. and {Tutuc}, Emanuel},
        title = "{Emergence of Interlayer Coherence in Twist-Controlled Graphene Double Layers}",
      journal = {Phys. Rev. Lett.},
         year = 2022,
        month = oct,
       volume = {129},
       number = {18},
          eid = {187701},
        pages = {187701},
          doi = {10.1103/PhysRevLett.129.187701},
archivePrefix = {arXiv},
       eprint = {2206.11799},
 primaryClass = {cond-mat.mes-hall},
       url = {https://link.aps.org/doi/10.1103/PhysRevLett.129.187701}
}

@article{qhf_collec_mode_Spielman_2001,
  title = {Observation of a Linearly Dispersing Collective Mode in a Quantum Hall Ferromagnet},
  author = {Spielman, I. B. and Eisenstein, J. P. and Pfeiffer, L. N. and West, K. W.},
  journal = {Phys. Rev. Lett.},
  volume = {87},
  issue = {3},
  pages = {036803},
  numpages = {4},
  year = {2001},
  month = {Jul},
  publisher = {American Physical Society},
  doi = {10.1103/PhysRevLett.87.036803},
  url = {https://link.aps.org/doi/10.1103/PhysRevLett.87.036803}
}

@article{quant_ferro_qhs_Yang_1994,
  title = {Quantum ferromagnetism and phase transitions in double-layer quantum Hall systems},
  author = {Yang, Kun and Moon, K. and Zheng, L. and MacDonald, A. H. and Girvin, S. M. and Yoshioka, D. and Zhang, Shou-Cheng},
  journal = {Phys. Rev. Lett.},
  volume = {72},
  issue = {5},
  pages = {732--735},
  numpages = {0},
  year = {1994},
  month = {Jan},
  publisher = {American Physical Society},
  doi = {10.1103/PhysRevLett.72.732},
  url = {https://link.aps.org/doi/10.1103/PhysRevLett.72.732}
}

@article{int_coh_qhs_merons_Moon_1995,
  title = {Spontaneous interlayer coherence in double-layer quantum Hall systems: Charged vortices and Kosterlitz-Thouless phase transitions},
  author = {Moon, K. and Mori, H. and Yang, Kun and Girvin, S. M. and MacDonald, A. H. and Zheng, L. and Yoshioka, D. and Zhang, Shou-Cheng},
  journal = {Phys. Rev. B},
  volume = {51},
  issue = {8},
  pages = {5138--5170},
  numpages = {0},
  year = {1995},
  month = {Feb},
  publisher = {American Physical Society},
  doi = {10.1103/PhysRevB.51.5138},
  url = {https://link.aps.org/doi/10.1103/PhysRevB.51.5138}
}

@article{GeickhBNEpsilonParallel1966,
  title = {Normal Modes in Hexagonal Boron Nitride},
  author = {Geick, R. and Perry, C. H. and Rupprecht, G.},
  journal = {Phys. Rev.},
  volume = {146},
  issue = {2},
  pages = {543--547},
  numpages = {0},
  year = {1966},
  month = {Jun},
  publisher = {American Physical Society},
  doi = {10.1103/PhysRev.146.543},
  url = {https://link.aps.org/doi/10.1103/PhysRev.146.543}
}

@article{YuEliashBNEpsilonParallel2013,
author = {G. L. Yu  and R. Jalil  and Branson Belle  and Alexander S. Mayorov  and Peter Blake  and Frederick Schedin  and Sergey V. Morozov  and Leonid A. Ponomarenko  and F. Chiappini  and S. Wiedmann  and Uli Zeitler  and Mikhail I. Katsnelson  and A. K. Geim  and Kostya S. Novoselov  and Daniel C. Elias },
title = {Interaction phenomena in graphene seen through quantum capacitance},
journal = {Proceedings of the National Academy of Sciences},
volume = {110},
number = {9},
pages = {3282-3286},
year = {2013},
doi = {10.1073/pnas.1300599110},
URL = {https://www.pnas.org/doi/abs/10.1073/pnas.1300599110},
}

@Article{LaturiahBNEpsilonParallel2018,
author={Laturia, Akash
and Van de Put, Maarten L.
and Vandenberghe, William G.},
title={Dielectric properties of hexagonal boron nitride and transition metal dichalcogenides: from monolayer to bulk},
journal={npj 2D Materials and Applications},
year={2018},
month={Mar},
day={08},
volume={2},
number={1},
pages={6},
issn={2397-7132},
doi={10.1038/s41699-018-0050-x},
url={https://doi.org/10.1038/s41699-018-0050-x}
}

@article{LeehBNEpsilonParallelQTM2026,
author = {Lee, M. and Das, I. and Herzog-Arbeitman, J. and Papp, J. and Li, J. and Daschner, M. and Zhou, Z. and Bhatt, M. and Currle, M. and Yu, J. and Jiang, Y. and Becherer, M. and Mittermeier, R. and Altpeter, P. and Obermayer, C. and Lorenz, H. and Chavez, G. and Le, B. T. and Williams, J. and Watanabe, K. and Taniguchi, T. and Bernevig, B. A. and Efetov, D. K.},
title = {Revealing Electron–Electron Interactions in Graphene at Room Temperature with a Quantum Twisting Microscope},
journal = {Nano Letters},
volume = {26},
number = {12},
pages = {4046-4052},
year = {2026},
doi = {10.1021/acs.nanolett.5c05015},
URL = {https://doi.org/10.1021/acs.nanolett.5c05015
}
}

@ARTICLE{chargeTranfer_Katayama_1995,
       author = {{Katayama}, Y. and {Tsui}, D.~C. and {Manoharan}, H.~C. and {Parihar}, S. and {Shayegan}, M.},
        title = "{Charge transfer at double-layer to single-layer transition in double-quantum-well systems}",
      journal = {Phys. Rev. B},
         year = 1995,
        month = nov,
       volume = {52},
       number = {20},
        pages = {14817-14824},
          doi = {10.1103/PhysRevB.52.14817},
          url = {https://link.aps.org/doi/10.1103/PhysRevB.52.14817}, 
       adsurl = {https://ui.adsabs.harvard.edu/abs/1995PhRvB..5214817K}
}

@ARTICLE{chargeTransfer_Manoharan_1997,
       author = {{Manoharan}, H.~C. and {Suen}, Y.~W. and {Lay}, T.~S. and {Santos}, M.~B. and {Shayegan}, M.},
        title = "{Spontaneous Interlayer Charge Transfer near the Magnetic Quantum Limit}",
      journal = {Phys. Rev. Lett.},
         year = 1997,
        month = oct,
       volume = {79},
       number = {14},
        pages = {2722-2725},
          doi = {10.1103/PhysRevLett.79.2722},
archivePrefix = {arXiv},
       eprint = {cond-mat/9709288},
 primaryClass = {cond-mat.mes-hall},
          url = {https://link.aps.org/doi/10.1103/PhysRevLett.79.2722},
       adsurl = {https://ui.adsabs.harvard.edu/abs/1997PhRvL..79.2722M}
}

@article{JIALiExcitonSuperfluidDoubleBilayerGraphene2017,
  title={Excitonic Superfluid Phase in Double Bilayer Graphene},
  author={J. I. A. Li and T. Taniguchi and K. Watanabe and J. Hone and C. R. Dean},
  journal={Nature Phys},
  volume={13},
  pages={751--755},
  year={2017}, 
  URL={https://doi.org/10.1038/nphys4140}, 
  doi={10.1038/nphys4140}
}

@ARTICLE{moireButterflies,
       author = {{Bistritzer}, R. and {MacDonald}, A.~H.},
        title = "{Moir{\'e} butterflies in twisted bilayer graphene}",
      journal = {Phys. Rev. B},
         year = 2011,
        month = jul,
       volume = {84},
       number = {3},
          eid = {035440},
        pages = {035440},
          doi = {10.1103/PhysRevB.84.035440},
archivePrefix = {arXiv},
       eprint = {1101.2606},
 primaryClass = {cond-mat.mes-hall},
       adsurl = {https://ui.adsabs.harvard.edu/abs/2011PhRvB..84c5440B}, 
      url = {https://link.aps.org/doi/10.1103/PhysRevB.84.035440}
}

@ARTICLE{moireBands,
       author = {{Bistritzer}, Rafi and {MacDonald}, Allan H.},
        title = "{Moir{\'e} bands in twisted double-layer graphene}",
      journal = {Proceedings of the National Academy of Science},
         year = 2011,
        month = jul,
       volume = {108},
       number = {30},
        pages = {12233-12237},
          doi = {10.1073/pnas.1108174108},
archivePrefix = {arXiv},
       eprint = {1009.4203},
 primaryClass = {cond-mat.mes-hall},
       adsurl = {https://ui.adsabs.harvard.edu/abs/2011PNAS..10812233B}
}

@ARTICLE{transportTBG,
       author = {{Bistritzer}, R. and {MacDonald}, A.~H.},
        title = "{Transport between twisted graphene layers}",
      journal = {Phys. Rev. B},
         year = 2010,
        month = jun,
       volume = {81},
       number = {24},
          eid = {245412},
        pages = {245412},
          doi = {10.1103/PhysRevB.81.245412},
archivePrefix = {arXiv},
       eprint = {1002.2983},
 primaryClass = {cond-mat.mes-hall},
       adsurl = {https://ui.adsabs.harvard.edu/abs/2010PhRvB..81x5412B},
      url = {https://link.aps.org/doi/10.1103/PhysRevB.81.245412}
}

@article{XiaomengQuantumHallDragExcitonGraphene2017,
  title={Quantum Hall drag of exciton condensate in graphene},
  author={Xiaomeng Liu and Kenji Watanabe and Takashi Taniguchi and Bertrand I. Halperin and Philip Kim},
  journal={Nature Phys},
  volume={13},
  pages={746--750},
  year={2017}, 
  URL={https://doi.org/10.1038/nphys4116}, 
  doi={10.1038/nphys4116}
}

@Article{YCaoTBGCorrelatedStates2018,
    author={Cao, Yuan
    and Fatemi, Valla
    and Demir, Ahmet
    and Fang, Shiang
    and Tomarken, Spencer L.
    and Luo, Jason Y.
    and Sanchez-Yamagishi, Javier D.
    and Watanabe, Kenji
    and Taniguchi, Takashi
    and Kaxiras, Efthimios
    and Ashoori, Ray C.
    and Jarillo-Herrero, Pablo},
    title={Correlated insulator behaviour at half-filling in magic-angle graphene superlattices},
    journal={Nature},
    year={2018},
    month={Apr},
    day={01},
    volume={556},
    number={7699},
    pages={80-84},
    issn={1476-4687},
    doi={10.1038/nature26154},
    url={https://doi.org/10.1038/nature26154}
}

@article{WillBurgCorrelatedStatesTDBG2019,
    title = {Correlated Insulating States in Twisted Double Bilayer Graphene},
    author = {Burg, G. William and Zhu, Jihang and Taniguchi, Takashi and Watanabe, Kenji and MacDonald, Allan H. and Tutuc, Emanuel},
    journal = {Phys. Rev. Lett.},
    volume = {123},
    issue = {19},
    pages = {197702},
    numpages = {5},
    year = {2019},
    month = {Nov},
    publisher = {American Physical Society},
    doi = {10.1103/PhysRevLett.123.197702},
    url = {https://link.aps.org/doi/10.1103/PhysRevLett.123.197702}
}

@article{ShenCorrelatedStatesTDBG2020,
    author={Shen, Cheng
    and Chu, Yanbang
    and Wu, QuanSheng
    and Li, Na
    and Wang, Shuopei
    and Zhao, Yanchong
    and Tang, Jian
    and Liu, Jieying
    and Tian, Jinpeng
    and Watanabe, Kenji
    and Taniguchi, Takashi
    and Yang, Rong
    and Meng, Zi Yang
    and Shi, Dongxia
    and Yazyev, Oleg V.
    and Zhang, Guangyu},
    title={Correlated states in twisted double bilayer graphene},
    journal={Nature Physics},
    year={2020},
    month={May},
    day={01},
    volume={16},
    number={5},
    pages={520-525},
    issn={1745-2481},
    doi={10.1038/s41567-020-0825-9},
    url={https://doi.org/10.1038/s41567-020-0825-9}
}

@article{XiaomengCorrelatedStatesTDBG2020,
    author={Liu, Xiaomeng
    and Hao, Zeyu
    and Khalaf, Eslam
    and Lee, Jong Yeon
    and Ronen, Yuval
    and Yoo, Hyobin
    and Haei Najafabadi, Danial
    and Watanabe, Kenji
    and Taniguchi, Takashi
    and Vishwanath, Ashvin
    and Kim, Philip},
    title={Tunable spin-polarized correlated states in twisted double bilayer graphene},
    journal={Nature},
    year={2020},
    month={Jul},
    day={01},
    volume={583},
    number={7815},
    pages={221-225},
    issn={1476-4687},
    doi={10.1038/s41586-020-2458-7},
    url={https://doi.org/10.1038/s41586-020-2458-7}
}

@Article{YCaoSuperconductivityMATBG2018,
    author={Cao, Yuan
    and Fatemi, Valla
    and Fang, Shiang
    and Watanabe, Kenji
    and Taniguchi, Takashi
    and Kaxiras, Efthimios
    and Jarillo-Herrero, Pablo},
    title={Unconventional superconductivity in magic-angle graphene superlattices},
    journal={Nature},
    year={2018},
    month={Apr},
    day={01},
    volume={556},
    number={7699},
    pages={43-50},
    issn={1476-4687},
    doi={10.1038/nature26160},
    url={https://doi.org/10.1038/nature26160}
}

@Article{JaneParkSuperconductivityMATTG2021,
    author={Park, Jeong Min
    and Cao, Yuan
    and Watanabe, Kenji
    and Taniguchi, Takashi
    and Jarillo-Herrero, Pablo},
    title={Tunable strongly coupled superconductivity in magic-angle twisted trilayer graphene},
    journal={Nature},
    year={2021},
    month={Feb},
    day={01},
    volume={590},
    number={7845},
    pages={249-255},
    issn={1476-4687},
    doi={10.1038/s41586-021-03192-0},
    url={https://doi.org/10.1038/s41586-021-03192-0}
}

@Article{WillBurgATQGCorrelations2022,
    author={Burg, G. William
    and Khalaf, Eslam
    and Wang, Yimeng
    and Watanabe, Kenji
    and Taniguchi, Takashi
    and Tutuc, Emanuel},
    title={Emergence of correlations in alternating twist quadrilayer graphene},
    journal={Nature Materials},
    year={2022},
    month={Aug},
    day={01},
    volume={21},
    number={8},
    pages={884-889},
    doi={10.1038/s41563-022-01286-2},
    url={https://doi.org/10.1038/s41563-022-01286-2}
}

@Article{JaneParkMultilayerGrapheneMagicSuperconductivity2022,
    author={Park, Jeong Min
    and Cao, Yuan
    and Xia, Li-Qiao
    and Sun, Shuwen
    and Watanabe, Kenji
    and Taniguchi, Takashi
    and Jarillo-Herrero, Pablo},
    title={Robust superconductivity in magic-angle multilayer graphene family},
    journal={Nature Materials},
    year={2022},
    month={Aug},
    day={01},
    volume={21},
    number={8},
    pages={877-883},
    issn={1476-4660},
    doi={10.1038/s41563-022-01287-1},
    url={https://doi.org/10.1038/s41563-022-01287-1}
}

@article{YZhangMultilayerGrapheneSuperconductivity2022,
  title     = "Promotion of superconductivity in magic-angle graphene
               multilayers",
  author    = "Zhang, Yiran and Polski, Robert and Lewandowski, Cyprian and
               Thomson, Alex and Peng, Yang and Choi, Youngjoon and Kim,
               Hyunjin and Watanabe, Kenji and Taniguchi, Takashi and Alicea,
               Jason and von Oppen, Felix and Refael, Gil and Nadj-Perge,
               Stevan",
  journal   = "Science",
  publisher = "American Association for the Advancement of Science (AAAS)",
  volume    =  377,
  number    =  6614,
  pages     = "1538--1543",
  month     =  sep,
  year      =  2022,
  doi={10.1126/science.abn8585}, 
  url={https://doi.org/10.1126/science.abn8585}
}

@Article{HanCorrelatedInsulatorChernInsulatorRhombohedral2024,
author={Han, Tonghang
and Lu, Zhengguang
and Scuri, Giovanni
and Sung, Jiho
and Wang, Jue
and Han, Tianyi
and Watanabe, Kenji
and Taniguchi, Takashi
and Park, Hongkun
and Ju, Long},
title={Correlated insulator and Chern insulators in pentalayer rhombohedral-stacked graphene},
journal={Nature Nanotechnology},
year={2024},
month={Feb},
day={01},
volume={19},
number={2},
pages={181-187},
issn={1748-3395},
doi={10.1038/s41565-023-01520-1},
url={https://doi.org/10.1038/s41565-023-01520-1}
}

@Article{ZhouSuperconductivityRhombohedral2021,
author={Zhou, Haoxin
and Xie, Tian
and Taniguchi, Takashi
and Watanabe, Kenji
and Young, Andrea F.},
title={Superconductivity in rhombohedral trilayer graphene},
journal={Nature},
year={2021},
month={Oct},
day={01},
volume={598},
number={7881},
pages={434-438},
issn={1476-4687},
doi={10.1038/s41586-021-03926-0},
url={https://doi.org/10.1038/s41586-021-03926-0}
}

@Article{ZhouFerromagnetismRhombohedral2021,
author={Zhou, Haoxin
and Xie, Tian
and Ghazaryan, Areg
and Holder, Tobias
and Ehrets, James R.
and Spanton, Eric M.
and Taniguchi, Takashi
and Watanabe, Kenji
and Berg, Erez
and Serbyn, Maksym
and Young, Andrea F.},
title={Half- and quarter-metals in rhombohedral trilayer graphene},
journal={Nature},
year={2021},
month={Oct},
day={01},
volume={598},
number={7881},
pages={429-433},
issn={1476-4687},
doi={10.1038/s41586-021-03938-w},
url={https://doi.org/10.1038/s41586-021-03938-w}
}

@article{AaronSharpeFerromagnetismTBG2019,
author = {Aaron L. Sharpe  and Eli J. Fox  and Arthur W. Barnard  and Joe Finney  and Kenji Watanabe  and Takashi Taniguchi  and M. A. Kastner  and David Goldhaber-Gordon },
title = {Emergent ferromagnetism near three-quarters filling in twisted bilayer graphene},
journal = {Science},
volume = {365},
number = {6453},
pages = {605-608},
year = {2019},
doi = {10.1126/science.aaw3780},
URL = {https://www.science.org/doi/abs/10.1126/science.aaw3780},
}

@Article{WangCorrlatedStatesTwistedTMD2020,
author={Wang, Lei
and Shih, En-Min
and Ghiotto, Augusto
and Xian, Lede
and Rhodes, Daniel A.
and Tan, Cheng
and Claassen, Martin
and Kennes, Dante M.
and Bai, Yusong
and Kim, Bumho
and Watanabe, Kenji
and Taniguchi, Takashi
and Zhu, Xiaoyang
and Hone, James
and Rubio, Angel
and Pasupathy, Abhay N.
and Dean, Cory R.},
title={Correlated electronic phases in twisted bilayer transition metal dichalcogenides},
journal={Nature Materials},
year={2020},
month={Aug},
day={01},
volume={19},
number={8},
pages={861-866},
issn={1476-4660},
doi={10.1038/s41563-020-0708-6},
url={https://doi.org/10.1038/s41563-020-0708-6}
}

@Article{KinFaiMakTwistedWSe2Supercon2025,
author={Xia, Yiyu
and Han, Zhongdong
and Watanabe, Kenji
and Taniguchi, Takashi
and Shan, Jie
and Mak, Kin Fai},
title={Superconductivity in twisted bilayer WSe2},
journal={Nature},
year={2025},
month={Jan},
day={01},
volume={637},
number={8047},
pages={833-838},
issn={1476-4687},
doi={10.1038/s41586-024-08116-2},
url={https://doi.org/10.1038/s41586-024-08116-2}
}

@Article{CoryDeanTwistedWSe2Supercon2025,
author={Guo, Yinjie
and Pack, Jordan
and Swann, Joshua
and Holtzman, Luke
and Cothrine, Matthew
and Watanabe, Kenji
and Taniguchi, Takashi
and Mandrus, David G.
and Barmak, Katayun
and Hone, James
and Millis, Andrew J.
and Pasupathy, Abhay
and Dean, Cory R.},
title={Superconductivity in 5.0{\textdegree} twisted bilayer WSe2},
journal={Nature},
year={2025},
month={Jan},
day={01},
volume={637},
number={8047},
pages={839-845},
issn={1476-4687},
doi={10.1038/s41586-024-08381-1},
url={https://doi.org/10.1038/s41586-024-08381-1}
}

@Article{XieChernInsulatorMATBG2021,
author={Xie, Yonglong
and Pierce, Andrew T.
and Park, Jeong Min
and Parker, Daniel E.
and Khalaf, Eslam
and Ledwith, Patrick
and Cao, Yuan
and Lee, Seung Hwan
and Chen, Shaowen
and Forrester, Patrick R.
and Watanabe, Kenji
and Taniguchi, Takashi
and Vishwanath, Ashvin
and Jarillo-Herrero, Pablo
and Yacoby, Amir},
title={Fractional Chern insulators in magic-angle twisted bilayer graphene},
journal={Nature},
year={2021},
month={Dec},
day={01},
volume={600},
number={7889},
pages={439-443},
issn={1476-4687},
doi={10.1038/s41586-021-04002-3},
url={https://doi.org/10.1038/s41586-021-04002-3}
}

@article{SerlinQAHEMoire2020,
author = {M. Serlin  and C. L. Tschirhart  and H. Polshyn  and Y. Zhang  and J. Zhu  and K. Watanabe  and T. Taniguchi  and L. Balents  and A. F. Young },
title = {Intrinsic quantized anomalous Hall effect in a moiré heterostructure},
journal = {Science},
volume = {367},
number = {6480},
pages = {900-903},
year = {2020},
doi = {10.1126/science.aay5533},
URL = {https://www.science.org/doi/abs/10.1126/science.aay5533},
}

@article{SpielmanResonantlyEnhancedDoubleLayerQuantumHallFerromagnet2000,
  title = {Resonantly Enhanced Tunneling in a Double Layer Quantum Hall Ferromagnet},
  author = {Spielman, I. B. and Eisenstein, J. P. and Pfeiffer, L. N. and West, K. W.},
  journal = {Phys. Rev. Lett.},
  volume = {84},
  issue = {25},
  pages = {5808--5811},
  numpages = {0},
  year = {2000},
  month = {Jun},
  publisher = {American Physical Society},
  doi = {10.1103/PhysRevLett.84.5808},
  url = {https://link.aps.org/doi/10.1103/PhysRevLett.84.5808}
}

@Article{NuckollsChernMATBG2020,
author={Nuckolls, Kevin P.
and Oh, Myungchul
and Wong, Dillon
and Lian, Biao
and Watanabe, Kenji
and Taniguchi, Takashi
and Bernevig, B. Andrei
and Yazdani, Ali},
title={Strongly correlated Chern insulators in magic-angle twisted bilayer graphene},
journal={Nature},
year={2020},
month={Dec},
day={01},
volume={588},
number={7839},
pages={610-615},
issn={1476-4687},
doi={10.1038/s41586-020-3028-8},
url={https://doi.org/10.1038/s41586-020-3028-8}
}

@Article{WuChernMATBG2021,
author={Wu, Shuang
and Zhang, Zhenyuan
and Watanabe, K.
and Taniguchi, T.
and Andrei, Eva Y.},
title={Chern insulators, van Hove singularities and topological flat bands in magic-angle twisted bilayer graphene},
journal={Nature Materials},
year={2021},
month={Apr},
day={01},
volume={20},
number={4},
pages={488-494},
issn={1476-4660},
doi={10.1038/s41563-020-00911-2},
url={https://doi.org/10.1038/s41563-020-00911-2}
}

@Article{YimengBulkEdgeTDBG2022,
author={Wang, Yimeng
and Herzog-Arbeitman, Jonah
and Burg, G. William
and Zhu, Jihang
and Watanabe, Kenji
and Taniguchi, Takashi
and MacDonald, Allan H.
and Bernevig, B. Andrei
and Tutuc, Emanuel},
title={Bulk and edge properties of twisted double bilayer graphene},
journal={Nature Physics},
year={2022},
month={Jan},
day={01},
volume={18},
number={1},
pages={48-53},
doi={10.1038/s41567-021-01419-5},
url={https://doi.org/10.1038/s41567-021-01419-5}
}

@article{TsuiFracQuanHallEffect1982,
  title = {Two-Dimensional Magnetotransport in the Extreme Quantum Limit},
  author = {Tsui, D. C. and Stormer, H. L. and Gossard, A. C.},
  journal = {Phys. Rev. Lett.},
  volume = {48},
  issue = {22},
  pages = {1559--1562},
  numpages = {0},
  year = {1982},
  month = {May},
  publisher = {American Physical Society},
  doi = {10.1103/PhysRevLett.48.1559},
  url = {https://link.aps.org/doi/10.1103/PhysRevLett.48.1559}
}

@article{LaughlinFracQuantHallEff1983,
  title = {Anomalous Quantum Hall Effect: An Incompressible Quantum Fluid with Fractionally Charged Excitations},
  author = {Laughlin, R. B.},
  journal = {Phys. Rev. Lett.},
  volume = {50},
  issue = {18},
  pages = {1395--1398},
  numpages = {0},
  year = {1983},
  month = {May},
  publisher = {American Physical Society},
  doi = {10.1103/PhysRevLett.50.1395},
  url = {https://link.aps.org/doi/10.1103/PhysRevLett.50.1395}
}

@Article{PKimFractQHEinGraphene2009,
author={Bolotin, Kirill I.
and Ghahari, Fereshte
and Shulman, Michael D.
and Stormer, Horst L.
and Kim, Philip},
title={Observation of the fractional quantum Hall effect in graphene},
journal={Nature},
year={2009},
month={Nov},
day={01},
volume={462},
number={7270},
pages={196-199},
doi={10.1038/nature08582},
url={https://doi.org/10.1038/nature08582}
}

@Article{CoryDeanGrapheneFracQHE2011,
author={Dean, C. R.
and Young, A. F.
and Cadden-Zimansky, P.
and Wang, L.
and Ren, H.
and Watanabe, K.
and Taniguchi, T.
and Kim, P.
and Hone, J.
and Shepard, K. L.},
title={Multicomponent fractional quantum Hall effect in graphene},
journal={Nature Physics},
year={2011},
month={Sep},
day={01},
volume={7},
number={9},
pages={693-696},
issn={1745-2481},
doi={10.1038/nphys2007},
url={https://doi.org/10.1038/nphys2007}
}

@article{GrimesWignerCrystal1979,
  title = {Evidence for a Liquid-to-Crystal Phase Transition in a Classical, Two-Dimensional Sheet of Electrons},
  author = {Grimes, C. C. and Adams, G.},
  journal = {Phys. Rev. Lett.},
  volume = {42},
  issue = {12},
  pages = {795--798},
  numpages = {0},
  year = {1979},
  month = {Mar},
  publisher = {American Physical Society},
  doi = {10.1103/PhysRevLett.42.795},
  url = {https://link.aps.org/doi/10.1103/PhysRevLett.42.795}
}

@article{EYAndreiWignerCrystal1988,
  title = {Observation of a Magnetically Induced Wigner Solid},
  author = {Andrei, E. Y. and Deville, G. and Glattli, D. C. and Williams, F. I. B. and Paris, E. and Etienne, B.},
  journal = {Phys. Rev. Lett.},
  volume = {60},
  issue = {26},
  pages = {2765--2768},
  numpages = {0},
  year = {1988},
  month = {Jun},
  publisher = {American Physical Society},
  doi = {10.1103/PhysRevLett.60.2765},
  url = {https://link.aps.org/doi/10.1103/PhysRevLett.60.2765}
}

@article{SantosWignerCrystal1992,
  title = {Observation of a reentrant insulating phase near the 1/3 fractional quantum Hall liquid in a two-dimensional hole system},
  author = {Santos, M. B. and Suen, Y. W. and Shayegan, M. and Li, Y. P. and Engel, L. W. and Tsui, D. C.},
  journal = {Phys. Rev. Lett.},
  volume = {68},
  issue = {8},
  pages = {1188--1191},
  numpages = {0},
  year = {1992},
  month = {Feb},
  publisher = {American Physical Society},
  doi = {10.1103/PhysRevLett.68.1188},
  url = {https://link.aps.org/doi/10.1103/PhysRevLett.68.1188}
}

@article{MoessnerStripeBubble1996,
  title = {Exact results for interacting electrons in high Landau levels},
  author = {Moessner, R. and Chalker, J. T.},
  journal = {Phys. Rev. B},
  volume = {54},
  issue = {7},
  pages = {5006--5015},
  numpages = {0},
  year = {1996},
  month = {Aug},
  publisher = {American Physical Society},
  doi = {10.1103/PhysRevB.54.5006},
  url = {https://link.aps.org/doi/10.1103/PhysRevB.54.5006}
}

@article{PanStripeBubble1999,
  title = {Strongly Anisotropic Electronic Transport at Landau Level Filling Factor $\mathit{\ensuremath{\nu}}\phantom{\rule{0ex}{0ex}}=\phantom{\rule{0ex}{0ex}}9/2$ and $\mathit{\ensuremath{\nu}}\phantom{\rule{0ex}{0ex}}=\phantom{\rule{0ex}{0ex}}5/2$ under a Tilted Magnetic Field},
  author = {Pan, W. and Du, R. R. and Stormer, H. L. and Tsui, D. C. and Pfeiffer, L. N. and Baldwin, K. W. and West, K. W.},
  journal = {Phys. Rev. Lett.},
  volume = {83},
  issue = {4},
  pages = {820--823},
  numpages = {0},
  year = {1999},
  month = {Jul},
  publisher = {American Physical Society},
  doi = {10.1103/PhysRevLett.83.820},
  url = {https://link.aps.org/doi/10.1103/PhysRevLett.83.820}
}

@article{LillyStripeBubble1999,
  title = {Evidence for an Anisotropic State of Two-Dimensional Electrons in High Landau Levels},
  author = {Lilly, M. P. and Cooper, K. B. and Eisenstein, J. P. and Pfeiffer, L. N. and West, K. W.},
  journal = {Phys. Rev. Lett.},
  volume = {82},
  issue = {2},
  pages = {394--397},
  numpages = {0},
  year = {1999},
  month = {Jan},
  publisher = {American Physical Society},
  doi = {10.1103/PhysRevLett.82.394},
  url = {https://link.aps.org/doi/10.1103/PhysRevLett.82.394}
}

@article{StanescuStripeBubble2000,
  title = {Finite-Temperature Density Instability at High Landau Level Occupancy},
  author = {Stanescu, Tudor D. and Martin, Ivar and Phillips, Philip},
  journal = {Phys. Rev. Lett.},
  volume = {84},
  issue = {6},
  pages = {1288--1291},
  numpages = {0},
  year = {2000},
  month = {Feb},
  publisher = {American Physical Society},
  doi = {10.1103/PhysRevLett.84.1288},
  url = {https://link.aps.org/doi/10.1103/PhysRevLett.84.1288}
}

@article{hBNSubstrates2010,
  title={Boron nitride substrates for high-quality graphene electronics},
  author={C. R. Dean and A. F. Young and I. Meric and C. Lee and L. Wang and S. Sorgenfrei and K. Watanabe and T. Taniguchi and P. Kim and K. L. Shepard and J. Hone},
  journal={Nature Nanotech},
  volume={5},
  pages={722--726},
  year={2010},
  URL={https://doi.org/10.1038/nnano.2010.172}, 
  doi={10.1038/nnano.2010.172}
}

@article{NitinQuantumLifeSpectroscopy2021,
  title = {Quantum Lifetime Spectroscopy and Magnetotunneling in Double Bilayer Graphene Heterostructures},
  author = {Prasad, Nitin and Burg, G. William and Watanabe, Kenji and Taniguchi, Takashi and Register, Leonard F. and Tutuc, Emanuel},
  journal = {Phys. Rev. Lett.},
  volume = {127},
  issue = {11},
  pages = {117701},
  numpages = {5},
  year = {2021},
  month = {Sep},
  publisher = {American Physical Society},
  doi = {10.1103/PhysRevLett.127.117701},
  url = {https://link.aps.org/doi/10.1103/PhysRevLett.127.117701}
}

@article{ManchesterTwistControlled2014,
  title={Twist-controlled resonant tunnelling in graphene/boron nitride/graphene heterostructures},
  author={A. Mishchenko and J. S. Tu and Y. Cao and R. V. Gorbachev and J. R. Wallbank and M. T. Greenaway and V. E. Morozov and S. V. Morozov and M. J. Zhu and S. L. Wong and F. Withers and C. R. Woods and Y-J. Kim and K. Watanabe and T. Taniguchi and E. E. Vdovin and O. Makarovsky and T. M. Fromhold and V. I. Falko and A. K. Geim and L. Eaves and K. S. Novoselov},
  journal={Nature Nanotech},
  volume={9},
  pages={808--813},
  year={2014},
  URL={https://doi.org/10.1038/nnano.2014.187},
  doi={10.1038/nnano.2014.187}
}

@article{ZhengTunnelingParallelTwoDim1993,
  title = {Tunneling conductance between parallel two-dimensional electron systems},
  author = {Zheng, Lian and MacDonald, A. H.},
  journal = {Phys. Rev. B},
  volume = {47},
  issue = {16},
  pages = {10619--10624},
  numpages = {0},
  year = {1993},
  month = {Apr},
  publisher = {American Physical Society},
  doi = {10.1103/PhysRevB.47.10619},
  url = {https://link.aps.org/doi/10.1103/PhysRevB.47.10619}
}

@article{FeenstraSingleParticleTunnelingGrapheneGraphene2012,
    author = {Feenstra, R. M. and Jena, Debdeep and Gu, Gong},
    title = {Single-particle tunneling in doped graphene-insulator-graphene junctions},
    journal = {Journal of Applied Physics},
    volume = {111},
    number = {4},
    pages = {043711},
    year = {2012},
    month = {02},
    issn = {0021-8979},
    doi = {10.1063/1.3686639},
    url = {https://doi.org/10.1063/1.3686639}
}

@article{SergioDeLaBerreraGrTunneling2014,
    author = {de la Barrera, Sergio C. and Gao, Qin and Feenstra, Randall M.},
    title = {Theory of graphene–insulator–graphene tunnel junctions},
    journal = {Journal of Vacuum Science and Technology B},
    volume = {32},
    number = {4},
    pages = {04E101},
    year = {2014},
    month = {04},
    issn = {2166-2746},
    doi = {10.1116/1.4871760},
    url = {https://doi.org/10.1116/1.4871760}
}

@article{WillDoubleBilayer2017,
author = {Burg, G. William and Prasad, Nitin and Fallahazad, Babak and Valsaraj, Amithraj and Kim, Kyounghwan and Taniguchi, Takashi and Watanabe, Kenji and Wang, Qingxiao and Kim, Moon J. and Register, Leonard F. and Tutuc, Emanuel},
title = {Coherent Interlayer Tunneling and Negative Differential Resistance with High Current Density in Double Bilayer Graphene–WSe2 Heterostructures},
journal = {Nano Letters},
volume = {17},
number = {6},
pages = {3919-3925},
year = {2017},
doi = {10.1021/acs.nanolett.7b01505},
URL = {https://doi.org/10.1021/acs.nanolett.7b01505}
}

@article{KyoungHighRotationalAccuracy2016,
author = {Kim, Kyounghwan and Yankowitz, Matthew and Fallahazad, Babak and Kang, Sangwoo and Movva, Hema C. P. and Huang, Shengqiang and Larentis, Stefano and Corbet, Chris M. and Taniguchi, Takashi and Watanabe, Kenji and Banerjee, Sanjay K. and LeRoy, Brian J. and Tutuc, Emanuel},
title = {van der Waals Heterostructures with High Accuracy Rotational Alignment},
journal = {Nano Letters},
volume = {16},
number = {3},
pages = {1989-1995},
year = {2016},
doi = {10.1021/acs.nanolett.5b05263},
URL = {https://doi.org/10.1021/acs.nanolett.5b05263}
}

@article{OneDimensionalEdgeContact2013,
author = {L. Wang  and I. Meric  and P. Y. Huang  and Q. Gao  and Y. Gao  and H. Tran  and T. Taniguchi  and K. Watanabe  and L. M. Campos  and D. A. Muller  and J. Guo  and P. Kim  and J. Hone  and K. L. Shepard  and C. R. Dean },
title = {One-Dimensional Electrical Contact to a Two-Dimensional Material},
journal = {Science},
volume = {342},
number = {6158},
pages = {614-617},
year = {2013},
doi = {10.1126/science.1244358},
URL = {https://www.science.org/doi/abs/10.1126/science.1244358},
eprint = {https://www.science.org/doi/pdf/10.1126/science.1244358},
}

@article{LyonGrapheneGS2017,
  title = {Probing Electron Spin Resonance in Monolayer Graphene},
  author = {Lyon, T. J. and Sichau, J. and Dorn, A. and Centeno, A. and Pesquera, A. and Zurutuza, A. and Blick, R. H.},
  journal = {Phys. Rev. Lett.},
  volume = {119},
  issue = {6},
  pages = {066802},
  numpages = {5},
  year = {2017},
  month = {Aug},
  publisher = {American Physical Society},
  doi = {10.1103/PhysRevLett.119.066802},
  url = {https://link.aps.org/doi/10.1103/PhysRevLett.119.066802}
}

@article{EichGrapheneGV2018,
  title = {Spin and Valley States in Gate-Defined Bilayer Graphene Quantum Dots},
  author = {Eich, Marius and Herman, Franti\ifmmode \check{s}\else \v{s}\fi{}ek and Pisoni, Riccardo and Overweg, Hiske and Kurzmann, Annika and Lee, Yongjin and Rickhaus, Peter and Watanabe, Kenji and Taniguchi, Takashi and Sigrist, Manfred and Ihn, Thomas and Ensslin, Klaus},
  journal = {Phys. Rev. X},
  volume = {8},
  issue = {3},
  pages = {031023},
  numpages = {11},
  year = {2018},
  month = {Jul},
  publisher = {American Physical Society},
  doi = {10.1103/PhysRevX.8.031023},
  url = {https://link.aps.org/doi/10.1103/PhysRevX.8.031023}
}

@article{LiSiYuGrapheneGV2019,
  title = {Scanning tunneling microscope study of quantum Hall isospin ferromagnetic states in the zero Landau level in a graphene monolayer},
  author = {Li, Si-Yu and Zhang, Yu and Yin, Long-Jing and He, Lin},
  journal = {Phys. Rev. B},
  volume = {100},
  issue = {8},
  pages = {085437},
  numpages = {6},
  year = {2019},
  month = {Aug},
  publisher = {American Physical Society},
  doi = {10.1103/PhysRevB.100.085437},
  url = {https://link.aps.org/doi/10.1103/PhysRevB.100.085437}
}

@Article{BanszerusGrapheneGV2021,
author={Banszerus, L.
and M{\"o}ller, S.
and Steiner, C.
and Icking, E.
and Trellenkamp, S.
and Lentz, F.
and Watanabe, K.
and Taniguchi, T.
and Volk, C.
and Stampfer, C.},
title={Spin-valley coupling in single-electron bilayer graphene quantum dots},
journal={Nature Communications},
year={2021},
month={Sep},
day={02},
volume={12},
number={1},
pages={5250},
doi={10.1038/s41467-021-25498-3},
url={https://doi.org/10.1038/s41467-021-25498-3}
}

\end{document}